\documentclass[a4paper,11pt]{article}
\pdfoutput=1 % if your are submitting a pdflatex (i.e. if you have
\usepackage{jheppub} % for details on the use of the package, please
\usepackage[T1]{fontenc} % if needed
\usepackage{braket}
\usepackage{upgreek}
\usepackage{subfigure}
\usepackage{xcolor}
\usepackage{bm}
\usepackage{multirow}
\usepackage{amsmath}
\DeclareMathOperator{\Tr}{Tr}
\usepackage{amsfonts}
\usepackage{bbold}

\def\arraystretch{1}

\everymath{\displaystyle}

\title{\boldmath A lattice study of $\pi\pi$ scattering at large $N_\text{c}$}

\author[a,1]{Jorge Baeza-Ballesteros\note{corresponding author},}
\author[a]{Pilar Hern\'andez,}
\author[a,b]{and Fernando Romero-L\'opez}

\affiliation[a]{IFIC, CSIC-Universitat de Val\`encia, 46980 Paterna, Spain}
\affiliation[b]{CTP, Massachusetts Institute of Technology, Cambridge, MA 02139, USA}

\emailAdd{jorge.baeza@uv.es}
\emailAdd{m.pilar.hernandez@uv.es}
\emailAdd{fernando@mit.edu}

\preprint{MIT-CTP/5395}

\abstract{ We present the first lattice study of pion-pion scattering with varying number of colors, $N_\text{c}$. We use lattice simulations with four degenerate quark flavors, $N_\text{f}=4$, and $N_\text{c}=3-6$. We focus on two scattering channels that do not involve vacuum diagrams. These correspond to two irreducible representations of the SU(4) flavor group: the fully symmetric one, $SS$, and the fully antisymmetric one, $AA$. The former is a repulsive channel equivalent to the isospin-2 channel of SU(2). By contrast, the latter is attractive and only exists for $N_\text{f} \geq 4$. A representative state is $\left( \ket{D_s^+ \pi^+} - \ket{D^+ K^+}\right)/\sqrt{2}$.  Using L\"uscher's formalism, we extract the near-threshold scattering amplitude
and we match our results to Chiral Perturbation Theory (ChPT) at large $N_\text{c}$. For this, we compute the analytical U$(N_\text{f})$ ChPT prediction for two-pion scattering, and use the lattice results to constrain the $N_\text{c}$ scaling of the relevant low-energy couplings.
}

\begin{document} 
\maketitle
\flushbottom

\newpage

\section{Introduction  \label{sec:intro}}
The 't Hooft limit of QCD~\cite{tHooft:1973alw} constitutes a significant simplification of the theory preserving all the interesting non-perturbative phenomena, such as asymptotic freedom, confinement and chiral symmetry breaking. In the limit of large number of colors, $N_{\text c}$, and a gauge coupling scaled as $g^2\propto N_{\text c}^{-1}$, QCD is believed to reduce to a theory of free, infinitely narrow and colorless resonances~\cite{tHooft:1973alw,Witten:1979kh,Coleman:1980nk} (i.e., glueballs and hadrons). 

Some controversy has arisen as regards the existence of exotic states, such as tetraquarks, in this limit. The  old standard lore concluded that tetraquarks cannot exist at large $N_\text{c}$~\cite{Coleman:1980nk}. However this reasoning has been revised recently and a different conclusion has been reached~\cite{PhysRevLett.110.261601,PhysRevD.88.036016}. The old argument simply implies that
the mixing of any tetraquark with a two-meson state vanishes in this limit, but says nothing on whether  tetraquarks actually exist. This is an interesting and timely question, in view of recent experimental discoveries of exotic states~\cite{Zyla:2020zbs}.

The description of the interactions and decays of these resonances requires non-vanishing $N^{-1}_{\text c}$ corrections.  A first-principle method that can quantify such corrections is the lattice formulation, where the number of colors can be varied~\cite{Teper:1998kw,Lucini:2001ej}. Indeed, various groups have successfully studied the $N_\text{c}$ scaling of some observables, such as the hadron or glueball spectrum, as well as matrix elements~\cite{Bali:2013kia,Cordon:2014sda,DeGrand:2016pur,DeGrand:2017gbi,DeGrand:2020utq,DeGrand:2021zjw,Athenodorou:2021qvs,Donini:2016lwz,Hernandez:2019qed,Donini:2020qfu,Perez:2020vbn}---see Ref.~\cite{Hernandez:2020tbc} for a recent review.

In this paper we present the first study of the $N_{\text c}$ scaling of the two-pion scattering amplitude using lattice QCD. We have used dynamical simulations at different number of colors in the range $N_{\text c}=3-6$ with $N_\text{f}=4$ degenerate quarks. This setup was used in Ref.~\cite{Donini:2020qfu} to study the $N_\text{c}$ scaling of the famous $\Delta I=1/2$ rule in the so-called GIM limit, where the charm is degenerate with the up quark. In this limit, penguin contractions vanish and therefore the enhancement of the isospin-0 over the isospin-2 channel depends solely on long-range QCD effects. The completion of that study requires the incorporation of final state interactions of the two pions, for which the study of $\pi\pi$ scattering is a first step.

We study the scattering in the two simplest channels, which do not involve vacuum contractions. They correspond to two irreducible representations (irreps) of SU(4). First, we have a fully symmetric 84-dimensional irrep, which is equivalent to the isospin-2 channel for $N_\text{f}=2$. Second, a 20-dimensional one which is antisymmetric for both quark and antiquark indices, contains states with four open flavors, and only exists for $N_\text{f} \geq 4$. Interestingly, as we will see, the antisymmetric channel turns out to be an attractive one, in contrast with the symmetric one which is repulsive.

We measure two-pion energy levels in lattice simulations from which we can extract infinite-volume scattering properties, such as the scattering length, using the celebrated L\"uscher's method~\cite{Luscher:1986pf,Luscher:1990ux}. The same observables can also be predicted in Chiral Perturbation Theory (ChPT)~\cite{Weinberg:1978kz,Gasser:1984gg}. These predictions can be found in the literature for $N_\text{c}=3$ and any $N_\text{f}$~\cite{Bijnens_2011}. However in the context of large $N_\text{c}$, it is necessary to include the $\eta'$ meson in the effective theory, since it becomes degenerate with the pions. The vacuum manifold is therefore not SU$(N_\text{f})$ but U$(N_\text{f})$~\cite{Kawarabayashi:1980dp,Herrera_Sikl_dy_1997,Kaiser_2000}. Somewhat surprisingly we have not found any computation of pion scattering in U$(N_\text{f})$ ChPT in the literature, so we present it here for the first time.

The comparison of lattice results with ChPT predictions can be used to determine the relevant low-energy couplings (LECs) of the chiral Lagrangian, and study their scaling in $N_\text{c}$. These results can be compared with those from large $N_\text{c}$ inspired phenomenological approaches such as Resonant Chiral Theory~\cite{Ecker1989,Ledwig:2014}. On the other hand, unitarized Chiral Theory  has been used to predict the fate of resonances at large $N_\text{c}$ from low-energy dynamics~\cite{GomezNicola:2001as, Pelaez:2003dy}. We will briefly comment on this approach for the attractive channel. Other approaches based on dispersion relations can also be found in the literature \ref{Dai:2017uao, Dai:2018fmx}.

The paper is organized as follows. In Sec.~\ref{sec:ChPT}, we present the results for the scattering amplitudes in ChPT with and without the $\eta'$. In Sec.~\ref{sec:finitevolume}, we review the finite-volume formalism, and in Sec.~\ref{sec:setup} we present our lattice setup. The results for two-pion scattering are presented in Sec.~\ref{sec:scattlen}, while the fits to ChPT and a comparison with other approaches is included in Sec.~\ref{sec:fits}.  We conclude in Sec.~\ref{sec:conclusions}. The paper contains two appendices: in App.~\ref{app:scatteringamplitude} we quote the main results for the scattering amplitudes in  SU($N_\text{f}$) and U($N_\text{f}$) ChPT, while in App.~\ref{app:WilsonChPT} we discuss how ChPT can be extended to study discretization effects.

\section{Chiral Perturbation Theory Predictions} \label{sec:ChPT}

Chiral Perturbation Theory (ChPT) describes QCD at energies below the QCD scale, $\mathit{\Lambda}_\text{QCD}$, in terms of the lightest non-singlet multiplet of pseudoscalar mesons (an octet for $N_\text{f}=3$). These are the pseudo-Nambu-Goldstone bosons (pNGB) arising from spontaneous chiral symmetry breaking, 
\begin{equation}
\text{SU}(N_\text{f})_\text{L}\times \text{SU}(N_\text{f})_\text{R}\rightarrow \text{SU}(N_\text{f})_\text{V}.
\end{equation}
The theory contains a number of unknown couplings, the low-energy couplings (LECs), that need to be determined from experiment or from first principles, i.e., via matching to lattice QCD. Mesonic observables can be computed in terms of these parameters as a perturbative expansion according to the standard power counting~\cite{Weinberg:1978kz, Gasser:1984gg},
\begin{equation} \label{eq:SUcountingChPT}
%\mathcal{O}(\delta)\sim
\mathcal{O}(m_q)\sim\mathcal{O}(M_\pi^2)\sim\mathcal{O}(k^2),
\end{equation}
where $m_q$ is the quark mass, and $M_\pi$ and $k^2$ are the meson mass and momenta, respectively.

A subtle point of ChPT in the 't Hooft limit is the treatment of the flavor-singlet pseudoscalar meson---the $\eta'$. According to the Witten-Veneziano relation~\cite{Witten1979, Veneziano1979}, the mass of this particle receives a large  contribution from the U(1)$_\text{A}$  anomaly:\footnote{We implicitly assume $N_\text{f}$ degenerate quarks.}
\begin{equation}
M_{\eta'}^2 -  M_\pi^2=  M_0^2 \equiv \frac{2N_\text{f} \chi_\text{YM}}{F_\pi^2} , \label{eq:WittenVeneziano}
\end{equation}
being $\chi_\text{YM}$ the topological susceptibility of the pure Yang-Mills theory, and $F_\pi$ the pion decay constant.\footnote{We use the normalization for $F_\pi$ corresponding to $\sim93$ MeV in QCD.} Since the $\chi_\text{YM} \sim N_\text{c}^0$ and $F_\pi^2 \sim N_\text{c}$, this contribution is suppressed as $N_\text{c}^{-1}$. Thus, the large $N_\text{c}$ pattern of chiral symmetry breaking becomes instead
\begin{equation}
\text{U}(N_\text{f})_\text{L}\times \text{U}(N_\text{f})_\text{R}\rightarrow \text{U}(N_\text{f})_\text{V},
\end{equation}
and the $\eta'$ is a pNGB that needs to be included in the effective theory at low energies.

The required modification of ChPT, sometimes referred to as U($N_\text{f}$) ChPT, has been studied before~\cite{DiVecchia:1980yfw, PhysRevD.21.3388, Witten:1980sp, Kawarabayashi:1980dp, Herrera_Sikl_dy_1997, Kaiser_2000}. It requires a modified power counting scheme that includes the scaling of $N_\text{c}^{-1}$.
A consistent one has been shown~\cite{Herrera_Sikl_dy_1997, Kaiser_2000} to be:
 \begin{equation} \label{eq:UcountingChPT}
%\mathcal{O}(\delta)\sim
\mathcal{O}(m_q)\sim\mathcal{O}(M_\pi^2)\sim\mathcal{O}(k^2)\sim\mathcal{O}(N_\text{c}^{-1}).
\end{equation}
While several quantities have been computed in U($N_\text{f}$) ChPT, meson-meson scattering has not been studied before in this context to the best of our knowledge. We therefore present here the first computation of scattering amplitudes of non-singlet mesons in U($N_\text{ f}$) ChPT, which will be necessary for this work. 

In a general theory with $N_\text{f}$ active flavors, there are seven different two-meson scattering channels\footnote{These are reduced  to three for $N_\text{f}=2$  and six for $N_\text{f}=3$.} of  non-singlet mesons~\cite{Bijnens_2011}. They correspond to different irreducible representations (irreps) of SU($N_\text{f}$). In this work, we focus on two of them:
\begin{itemize}
\item The irrep of maximal dimensionality---84 for $N_\text{f}=4$---that is symmetric in both quarks and antiquarks. It is equivalent to the isospin-2 channel of SU(2). We will refer to it as $SS$ channel.  A representative state of this channel is the well-known $|\pi^\pm\pi^\pm\rangle$.
\item The irrep containing states with four distinct quark flavors that is antisymmetric in both quarks and antiquarks. This one only exists for $N_\text{f}\geq 4$, and is 20-dimensional for $N_\text{f}=4$. We denote it as the $AA$ channel. A representative state is $(\ket{D_s^+ \pi^+} - \ket{D^+ K^+})/{\sqrt{2}} $.
\end{itemize}
Throughout this work, we will generically refer to any of them as $R$.

%We now compare the predictions of the scattering amplitudes in these two channels for SU($N_\text{f}$) ChPT and U($N_\text{f}$) ChPT. 

\subsection{SU($N_\text{f}$) predictions}\label{sec:SUPredictions}

At leading order (LO) in the chiral expansion, the pion-pion scattering amplitude depends only on the ratio $M_\pi/F_\pi$. At this order the result is found to be the same for both channels up to a sign
\begin{eqnarray}
\mathcal{M}^{SS}_\text{LO}=-\mathcal{M}^{AA}_\text{LO} = \frac{M_\pi^2}{F_\pi^2}(2-s), \label{eq:MLO}
\end{eqnarray}
where $s$ is the usual Mandelstam variable normalized by the pion mass. 

At next-to-leading-order (NLO) and for $N_\text{f}\geq 4$ a total of 13 additional LECs become relevant in the chiral Lagrangian, albeit only some linear combinations appear in the observables of interest. The scattering amplitudes for the two channels are known up to next-to-next-to-leading-order (NNLO) for $N_\text{f}$ degenerate quark flavors~\cite{Bijnens_2011}.  For completeness we reproduce the $s$-wave projected NLO result  in App.~\ref{app:scatteringamplitude}. From these amplitudes, one can extract the scattering lengths which are found to be:\footnote{We use the convention in which $k\cot\delta_0=1/a_0$, where $a_0$ is the $s$-wave phase shift.}
\begin{multline}\label{eq:I2SUScatteringLength}
M_{\pi}a_0^{SS} =-\frac{M_{\pi}^2}{16\pi F_{\pi}^2}\left[1-\frac{16 M_{\pi}^2}{F_{\pi}^2}L_{SS}\right. +\frac{M_{\pi}^2}{8F_{\pi}^2\pi^2 N_{\text{f}}^2}\\\left.-\frac{M_{\pi}^2}{8F_{\pi}^2\pi^2 N_{\text{f}}}+\frac{M_{\pi}^2}{8F_{\pi}^2\pi^2}\log{\frac{M_{\pi}^2}{\mu^2}}+\frac{M_{\pi}^2}{8F_{\pi}^2\pi^2N_{\text{f}}^2}\log{\frac{M_{\pi}^2}{\mu^2}}-\frac{M_\pi^2}{8F_{\pi}^2\pi^2N_{\text{f}}}\log{\frac{M_{\pi}^2}{\mu^2}}\right],
\end{multline}
\begin{multline}\label{eq:AASUScatteringLength}
M_{\pi}a_0^{AA} =\frac{M_{\pi}^2}{16\pi F_{\pi}^2}\left[1-\frac{16 M_{\pi}^2}{F_{\pi}^2}L_{AA}\right. -\frac{M_{\pi}^2}{8F_{\pi}^2\pi^2 N_{\text{f}}^2}\\\left.-\frac{M_{\pi}^2}{8F_{\pi}^2\pi^2 N_{\text{f}}}-\frac{M_{\pi}^2}{8F_{\pi}^2\pi^2}\log{\frac{M_{\pi}^2}{\mu^2}}-\frac{M_{\pi}^2}{8F_{\pi}^2\pi^2N_{\text{f}}^2}\log{\frac{M_{\pi}^2}{\mu^2}}-\frac{M_\pi^2}{8F_{\pi}^2\pi^2N_{\text{f}}}\log{\frac{M_{\pi}^2}{\mu^2}}\right],
\end{multline}
where  $L_{R}\equiv L_R(\mu)$ are the following linear combinations of LECs,  defined at the renormalization scale $\mu$,
\begin{equation}\label{eq:I2LECs}
L_{SS} = L_0 + 2 L_1 + 2 L_2 + L_3 - 2 L_4 - L_5 + 2 L_6 + L_8,
\end{equation}
\begin{equation}\label{eq:AALECs}
L_{AA} = L_0 - 2 L_1 - 2 L_2 + L_3 +2 L_4 - L_5 - 2 L_6 + L_8.
\end{equation}
Eqs.~(\ref{eq:I2SUScatteringLength}) and (\ref{eq:AASUScatteringLength}) are valid for any number of colors, but the LECs, $F_\pi^2$ and $M_\pi^2$ have an implicit $N_\text{c}$ scaling~\cite{Manohar:1998xv}:
\begin{equation}\label{eq:LECscaling}
\mathcal{O}(N_\text{c}):F_\pi^2, L_0, L_3, L_5, L_8, \quad\quad\quad \mathcal{O}(1): M_\pi, L_1, L_2, L_4, L_6.
\end{equation}
The leading $N_\text{c}$ scaling of $L_R$ is thus the same, but subleading effects differ. We
therefore expect:
\begin{equation}
\def\arraystretch{1.2}
\begin{array}{rl}
L_{SS}&=N_\text{c}L^{(0)}+L^{(1)}_{SS}+\mathcal{O}(N_\text{c}^{-1}),\\
L_{AA}&=N_\text{c}L^{(0)}+L^{(1)}_{AA}+\mathcal{O}(N_\text{c}^{-1}). \label{eq:LECsExpansion}
\end{array}
\end{equation}

Another quantity of interest that may be extracted from the scattering amplitude is the effective range, $r_0$. At LO both channels have
\begin{equation}\label{eq:EffectiveRangeLO}
M_\pi^2a_0^{R}r_0^{R}=-3,
\end{equation}
meaning $r_0$ is positive (negative) in the $SS$ ($AA$) channel.

\subsection{U($N_\text{f}$) predictions}\label{sec:UPredictions}

We use the same notation for the LECs in the two chiral theories, even though they will take different values, as we will see when we discuss the matching between the two. 

At NLO in the U$(N_\text{f})$ power counting of Eq.~(\ref{eq:UcountingChPT}), the different observables are given by the tree-level contribution from the LO Lagrangian, and the NLO corrections from the LECs of order $O(N_\text{c})$. 
For instance, the $SS$ scattering length is given by
\begin{equation}
M_\pi a_0^{SS} =  \frac{M_{\pi}^2}{16\pi F_{\pi}^2}\left[1-\frac{16 M_{\pi}^2}{F_{\pi}^2} N_\text{c} L^{(0)} \right],
\end{equation}
and similarly for the $AA$ channel. This prediction is however not sufficient~\cite{Hernandez:2019qed}, and one needs to go to NNLO in the U$(N_\text{f})$ power counting. At this order, loop diagrams---and thus chiral logarithms---start to contribute, and further care has to be taken to include the $\eta'$ meson. Note that, we use the same notation for the LECs in the two chiral theories, even though they will take different values, as we will see when we discuss the matching between the two.

\begin{figure}[h!]
\centering

   \subfigure[]%
             {\includegraphics[scale=0.3]{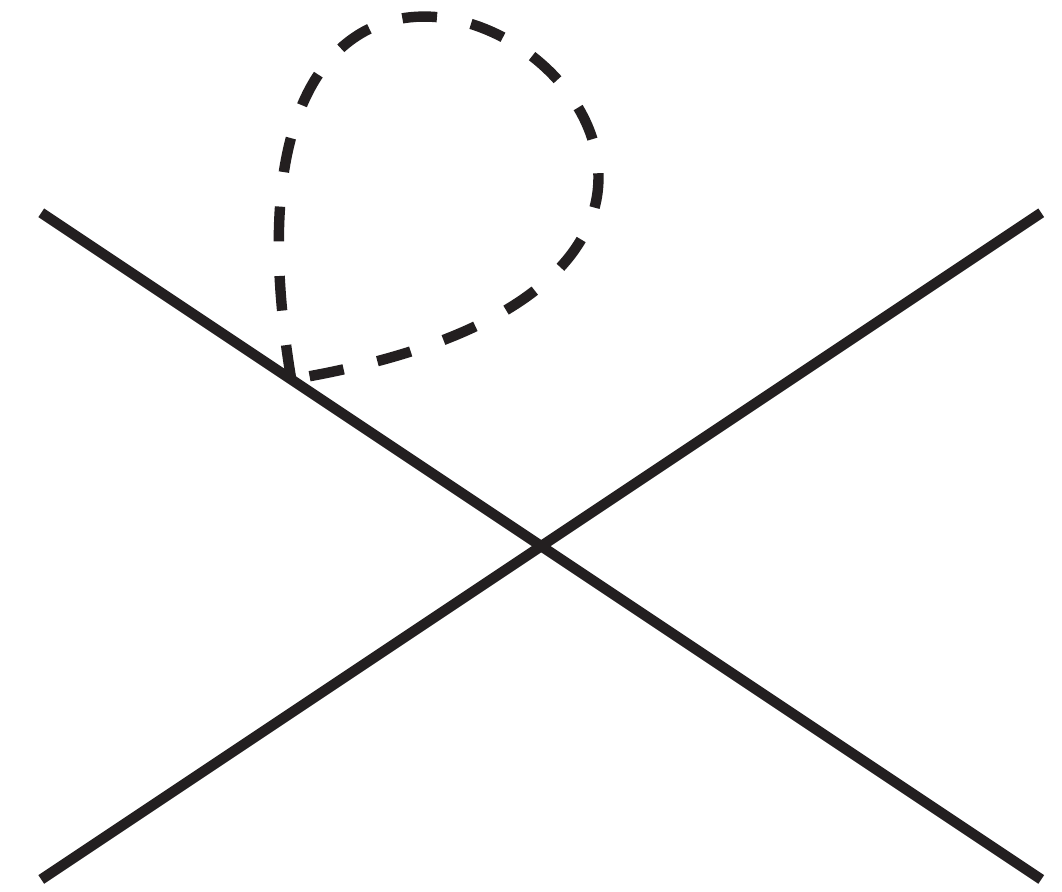} }
             \hspace{0.4cm}
   \subfigure[]%
             {\includegraphics[scale=0.3]{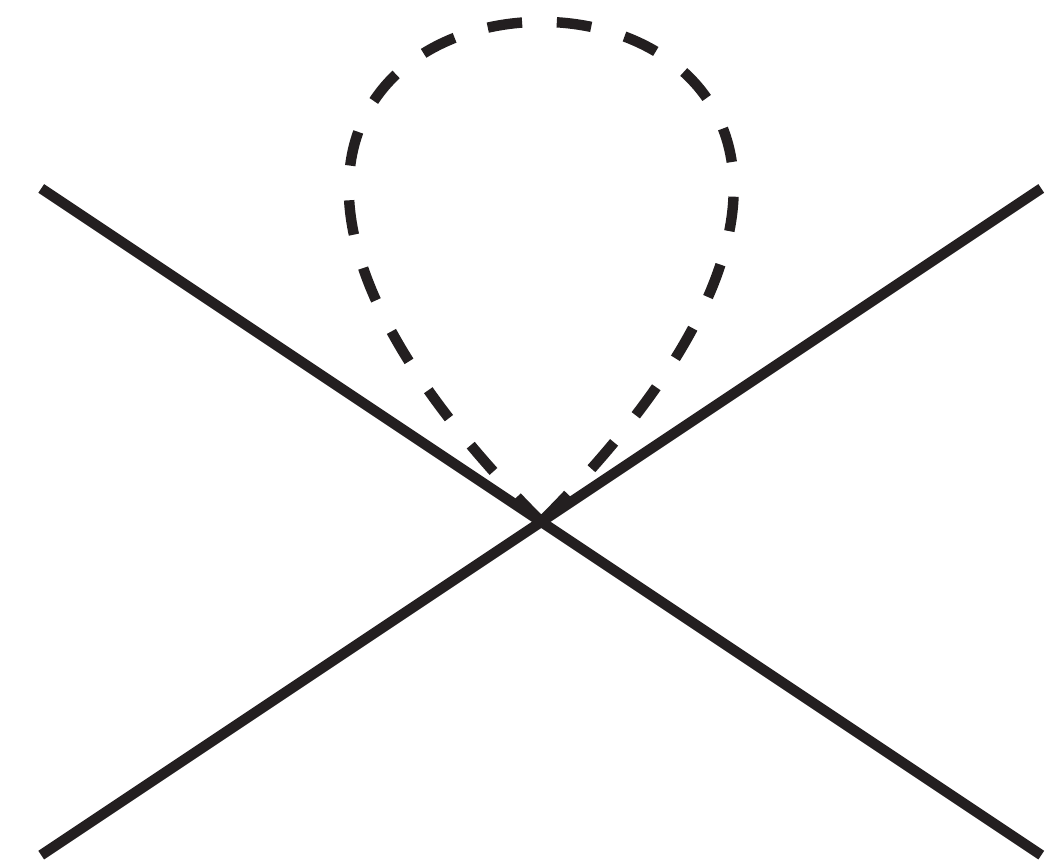} } \hspace{0.4cm}
   \subfigure[]%
             {\includegraphics[scale=0.3]{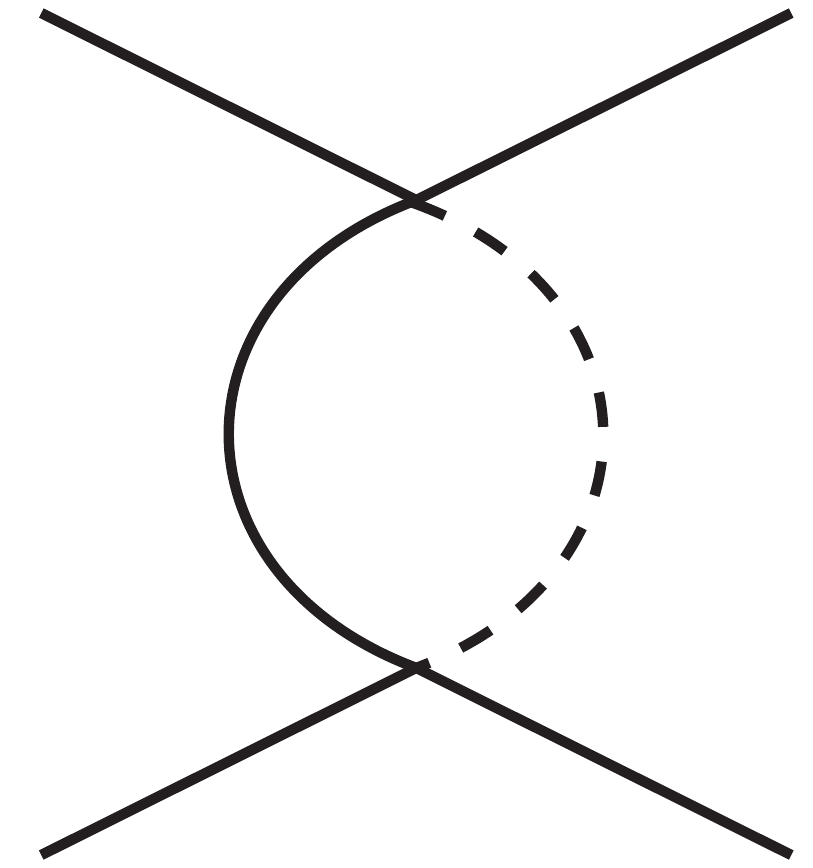}  \label{fig:loop1eta}}\hspace{0.4cm}
   \subfigure[]%
             {\includegraphics[scale=0.3]{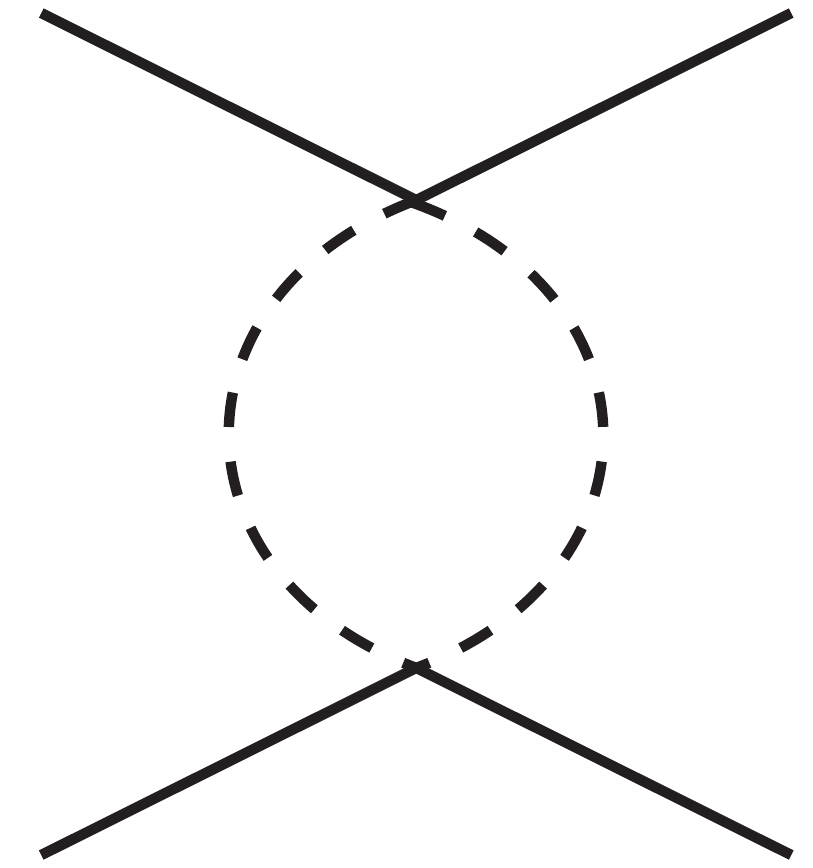}  \label{fig:loop2eta}}
\caption{~Additional one-loop Feynman diagrams required to study two-pion scattering for the two channels of interest in U($N_\text{f}$) ChPT. Solid lines depict non-singlet mesons, while dotted represent the $\eta'$.}\label{fig:SUversusU}
\end{figure}

The contribution of the $\eta'$ meson is encoded in a few additional diagrams with respect to the SU$(N_\text{f})$ result. These are summarized in Fig.~\ref{fig:SUversusU}, where solid (dashed) lines represent multiplet (singlet) mesons. The NNLO result is given by
\begin{equation}\label{eq:additionalUdiagrams}
\mathcal{M}^{\text{U}(N_\text{f})}_\text{NNLO}  =  \mathcal{M}^{\text{SU}(N_\text{f})}_\text{NLO} + \mathcal{M}_K  + \Delta Z^2_{\text{U}(N_\text{f})} \mathcal{M}_\text{LO}  + \mathcal{M}^\text{loop}_{\eta'},
\end{equation}
where $\Delta Z^2$ represents the additional mass renormalization due to $\eta'$ loops---diagram~\ref{fig:SUversusU}(a)---and $\mathcal{M}^\text{loop}_{\eta'}$ represents diagrams~\ref{fig:SUversusU}(b) --~\ref{fig:SUversusU}(d), that involve $\eta'$ loops. Finally, $\mathcal{M}_K$ includes products of $\mathcal{O}(N_\text{c})$ LECs.

Our results for the scattering amplitudes at NNLO are presented in App~\ref{app:scatteringamplitude}. From these we obtain the following  scattering lengths:
\begin{equation}\label{eq:I2UScatteringLength}
\def\arraystretch{2}
\begin{array}{rl}
M_\pi a_0^{SS}
=&-\frac{M_{\pi}^2}{16\pi F_{\pi}^2}\left[1-\frac{16 M_{\pi}^2}{F_{\pi}^2}L_{SS}+K_{SS}\left(\frac{M_\pi^2}{F_{\pi}^2}\right)^2 +\frac{M_{\pi}^2}{4F_{\pi}^2\pi^2 N_{\text{f}}^2} - \frac{M_{\pi}^2}{8F_{\pi}^2\pi^2 N_{\text{f}}}\right.\\ 
&\left.  + \frac{M_{\pi}^2}{8F_{\pi}^2\pi^2}\log{\frac{M_{\pi}^2}{\mu^2}} - \frac{M_\pi^2+ M_{\eta'}^2}{M_\pi^2-M_{\eta'}^2}\frac{M_{\pi}^2}{8F_{\pi}^2\pi^2N_{\text{f}}^2}\log{\frac{M_{\pi}^2}{\mu^2}}+ \frac{ M_{\eta'}^2}{M_\pi^2-M_{\eta'}^2}\frac{M_\pi^2}{8F_{\pi}^2\pi^2N_{\text{f}}}\log{\frac{M_{\pi}^2}{\mu^2}}\right.\\
&\left. + \frac{M_\pi^2+ M_{\eta'}^2}{M_\pi^2-M_{\eta'}^2}\frac{M_\pi^2}{8F_{\pi}^2\pi^2N_{\text{f}}^2}\log{\frac{M_{\eta'}^2}{\mu^2}} - \frac{ M_{\eta'}^2}{M_\pi^2-M_{\eta'}^2}\frac{M_\pi^2}{8F_{\pi}^2\pi^2N_{\text{f}}}\log{\frac{M_{\eta'}^2}{\mu^2}}\right],
\end{array}
\end{equation}
\begin{equation}
\label{eq:AAUScatteringLength}
\def\arraystretch{2}
\begin{array}{rl}
M_\pi a_0^{AA}
=&\frac{M_{\pi}^2}{16\pi F_{\pi}^2}\left[1-\frac{16 M_{\pi}^2}{F_{\pi}^2}L_{AA}+K_{AA}\left(\frac{M_\pi^2}{F_{\pi}^2}\right)^2 -\frac{M_{\pi}^2}{4F_{\pi}^2\pi^2 N_{\text{f}}^2} - \frac{M_{\pi}^2}{8F_{\pi}^2\pi^2 N_{\text{f}}}\right.\\ 
&\left.  - \frac{M_{\pi}^2}{8F_{\pi}^2\pi^2}\log{\frac{M_{\pi}^2}{\mu^2}} + \frac{M_\pi^2+ M_{\eta'}^2}{M_\pi^2-M_{\eta'}^2}\frac{M_{\pi}^2}{8F_{\pi}^2\pi^2N_{\text{f}}^2}\log{\frac{M_{\pi}^2}{\mu^2}}+ \frac{ M_{\eta'}^2}{M_\pi^2-M_{\eta'}^2}\frac{M_\pi^2}{8F_{\pi}^2\pi^2N_{\text{f}}}\log{\frac{M_{\pi}^2}{\mu^2}}\right.\\
&\left. - \frac{M_\pi^2+ M_{\eta'}^2}{M_\pi^2-M_{\eta'}^2}\frac{M_\pi^2}{8F_{\pi}^2\pi^2N_{\text{f}}^2}\log{\frac{M_{\eta'}^2}{\mu^2}} - \frac{ M_{\eta'}^2}{M_\pi^2-M_{\eta'}^2}\frac{M_\pi^2}{8F_{\pi}^2\pi^2N_{\text{f}}}\log{\frac{M_{\eta'}^2}{\mu^2}}\right],
\end{array}
\end{equation}
where  $M_{\eta'}$ is given by the Witten-Veneziano formula in Eq.~(\ref{eq:WittenVeneziano}). 
%Note that for degenerate quarks, there is no mixing between the $\eta'$ and other neutral mesons~\cite{Guo:2015xva}.   
$K_R$ are combinations of products of  LECs and are $\mathcal{O}(N_\text{c}^2)$.  From the results of Ref.~\cite{Bijnens_2011}, we find that the leading $N_\text{c}$ dependence of these factors is also equal for both channels, i.e.,
\begin{equation}\label{eq:leadingK}
 K_{SS}\Big\rvert_{\mathcal{O}(N_\text{c}^2)}=  K_{AA}\Big\rvert_{\mathcal{O}(N_\text{c}^2)}= 128(L_8-2L_5)^2\Big\rvert_{\mathcal{O}(N_\text{c}^2)} .
\end{equation}

It is also possible to get the prediction for the effective range in the U($N_\text{f}$) theory as was done for SU($N_\text{f}$) ChPT. The LO result is the same as in Eq.~(\ref{eq:EffectiveRangeLO}) while the NNLO result can be reconstructed from the scattering amplitudes in App.~\ref{app:scatteringamplitude}.

\subsection{Large $N_\text{c}$ dependence}

In the U$(N_\text{f})$ theory, the $N_\text{c}$ scaling of the LECs  in Eq.~(\ref{eq:LECsExpansion}) holds, with a common leading $N_\text{c}$ dependence for both channels. 
%We can therefore assume that the scaling of the NLO LECs must be of the form
%\begin{equation}
%L_R \Big\rvert_{\text{U}(N_\text{f})} = N_\text{c} L^{(0)} + L_R^{(1)}+\mathcal{O}(N_\text{c}^{-1}),
%\end{equation}
%where $R=AA$ or $SS$. Note that $L^{(0)}$ is common for both channels, while $L_R^{(1)}$ is not. 
It remains to be seen whether it is possible to derive a simple $N_\text{f}$ dependence for the $L_R^{(1)}$ terms in this theory,  by means of a perturbative analysis of correlation functions at large $N_\text{c}$.

\begin{figure}[h!]
\centering
   \subfigure[$D$ diagram, $\mathcal{O}(N_\text{c}^2)$]%
             {\includegraphics[width=0.35\textwidth]{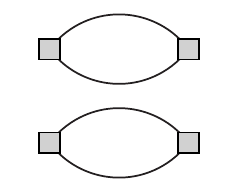} \label{fig:Dleading}}
   \subfigure[$C$ diagram, $\mathcal{O}(N_\text{c}^1)$]%
             {\includegraphics[width=0.35\textwidth]{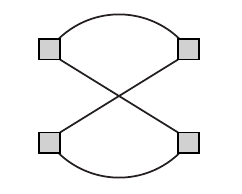} \label{fig:Cleading}} 
\caption{\label{fig:contractions} Diagrammatic representation of the quark  Wick contractions contributing to the $SS$ and $AA$ channels. Grey squares represent a pion operator and solid lines are quark lines.}
\end{figure}

The scattering of two pions in these channels at the quark level involves two topologies for the quark Wick contractions, as depicted in Fig.~\ref{fig:contractions}. We denote them as disconnected, $D$, and connected, $C$. Their contribution to the meson-meson correlator in the two channels is 
\begin{equation}
C_{SS}(t)=2(D-C), \quad\quad C_{AA}(t)=2(D+C). \label{eq:wickcontractions}
\end{equation}
There are two color loops in the  $D$ diagram, and just one in the $C$ diagram. Thus, they are $\mathcal{O}(N_\text{c}^2)$ and $\mathcal{O}(N_\text{c})$, respectively.

In terms of correlation functions, the scattering length is roughly given by:
\begin{equation}
a^{R}_0 \propto \frac{C_R - C_\pi^2}{C_\pi^2}, \label{eq:a0scaling}
\end{equation}
where $C_\pi\sim\mathcal{O}(N_\text{c})$ is the single-pion correlator, which contains a single color loop. One can see that factorizable diagrams do not contribute to the scattering length. For instance, the $N_\text{c}$-leading diagrams contributing to $D$ in Fig.~\ref{fig:Dleading} are planar contributions that involve gluons exchanged within each loop, and get exactly cancelled by $C_\pi^2$. 
\begin{figure}[h]
   \centering

   \subfigure[$\mathcal{O}(N_\text{c}^0)$]%
             {\includegraphics[width=0.35\textwidth]{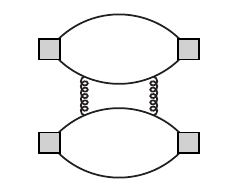}  \label{fig:Dsubleading}}
   \subfigure[$\mathcal{O}(N_\text{f} N_\text{c}^0)$]%
             {\includegraphics[width=0.33\textwidth]{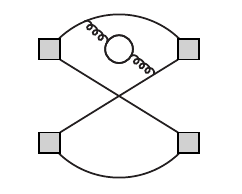}  \label{fig:Csubleading}}
   \caption{ Subleading diagrams in the large $N_\text{c}$ scaling of the Wick contractions $D$ and $C$ in Fig.~\ref{fig:contractions}. Grey squares represent a pion operator, solid lines represent quark lines, and
  wavy lines are gluons. The scaling of each diagram is given in their respective subcaptions.}
  \label{fig:diagslargeNc}
\end{figure}

We can now analyze the leading non-factorizable contributions of $D$ and $C$. The latter is non-factorizable and of $\mathcal{O}(N_\text{c})$ [see Fig.~\ref{fig:Cleading}] yielding the leading contribution to the scattering length. Exchanging planar gluons does not change the scaling, and including a quark loop contributes with an extra power of $N_\text{f}/N_\text{c}$ [Fig.~\ref{fig:Csubleading}]. Concerning the nonfactorizable contributions in $D$, they necessarily involve gluon exchange between the two loops, for example Fig.~\ref{fig:Dsubleading}. This diagram is suppressed as $1/N_\text{c}^2$ relative to the leading scaling of $D$, and is thus $\mathcal{O}(N_\text{c}^0)$. Finally, the scaling $C_\pi$  scales like $C$. All this can be summarized as follows:

\begin{align}
\begin{split}
C &= {N_\text{c}} \left(a + b \frac{N_\text{f}}{N_\text{c}} \right) +\mathcal{O}(N_\text{c}^{-1}), \\
D - C_\pi^2 &=  c +\mathcal{O}(N_\text{c}^{-1}), \\
C_\pi &= {N_\text{c}}\left(d + e \frac{N_\text{f}}{N_\text{c}} \right) +\mathcal{O}(N_\text{c}^{-1}),
\end{split}
\end{align}
where $a-e$ are numerical constants, independent of $N_\text{c}$ and $N_\text{f}$. Combining the scaling of the diagrams, with the linear combinations in Eq.~(\ref{eq:wickcontractions}), one can derive
\begin{equation}
M_\pi a^{R}_0 \approx \pm \frac{1}{N_\text{c}} \left( \tilde a + \tilde b \frac{N_\text{f}}{N_\text{c}} \mp \tilde c \frac{1}{N_\text{c}}   \right) +\mathcal{O}(N_\text{c}^{-3}), \label{eq:resa0scaling}
\end{equation}
where $\tilde a-\tilde c$ are linear combinations of $a-e$, and the upper (lower) signs correspond to the $AA$ ($SS$) channel.  large $N_\text{c}$ arguments therefore explain why the repulsive nature of one channel implies the attractive nature of the other.

By matching Eq.~(\ref{eq:resa0scaling}) to the ChPT predictions in Eqs.~(\ref{eq:I2UScatteringLength}) and (\ref{eq:AAUScatteringLength}), we can derive the first coefficients of the large $N_\text{c}$ scaling of the NLO LECs. We find
\begin{equation}
L_R \Big\rvert_{\text{U}(N_\text{f})} = N_\text{c} L^{(0)} + N_\text{f} L_\text{c}^{(1)} \mp L_\text{a}^{(1)} +\mathcal{O}(N_\text{c}^{-1}),
\label{eq:LRscaling}
\end{equation}
where the correlated and anticorrelated terms, $L_\text{c}^{(1)}$ and $L_\text{a}^{(1)}$, are now independent of $N_\text{f}$. Comparing to Eqs.~(\ref{eq:I2LECs}) and (\ref{eq:AALECs}), one can see that
\begin{align}
\begin{split}
L_0 + L_3 - L_5 + L_8 &= N_\text{c} L^{(0)} + N_\text{f} L_\text{c}^{(1)}  +\mathcal{O}(N_\text{c}^{-1}), \\
2 L_1 + 2 L_2 - 2 L_4 + 2 L_6 &= L_\text{a}^{(1)} +\mathcal{O}(N_\text{c}^{-1})
\end{split}
\end{align}
Finally, one may also check that the factors of $1/N_\text{f}$ in Eqs.~(\ref{eq:I2UScatteringLength}) and (\ref{eq:AAUScatteringLength}) explicitly cancel at large $N_\text{c}$. This can be seen analytically assuming the Witten-Veneziano relation for the $\eta'$ mass [Eq.~(\ref{eq:WittenVeneziano})], and expanding in $1/N_\text{c}$. In conclusion, the U($N_\text{f}$) ChPT predictions are found to be consistent with the diagramatic $1/N_\text{c}$ expansion in Eq.~(\ref{eq:resa0scaling}), with the $M_\pi$ dependence hiding in the values of the coefficients  $\tilde a-\tilde c$.

\subsection{Matching the U$(N_\text{f})$ and SU$(N_\text{f})$ theories}

The standard form of SU$(N_\text{f})$ ChPT can be recovered from the U($N_\text{f}$) one in the limit of large $\eta'$ mass, $M_{\eta'}\gg M_\pi$. The effect of integrating out this particle remains encoded in the SU($N_\text{f}$) LECs, and results in corrections that scale as $1/N_\text{f}$ and $1/N_\text{f}^2$. The matching between both theories  at NNLO gives the following relations:
\begin{equation}\label{eq:I2L1Matching}
L^{(1)}_{SS}\Big\rvert_{\text{SU}(N_\text{f})}=L^{(1)}_{SS}\Big\rvert_{\text{U}(N_\text{f})}-\frac{1}{8N_{\text{f}}^2(4\pi)^2}(N_\text{f}\lambda_0-\lambda_0+1),
\end{equation}
\begin{equation}\label{eq:AAL1Matching}
L^{(1)}_{AA}\Big\rvert_{\text{SU}(N_\text{f})}=L^{(1)}_{AA}\Big\rvert_{\text{U}(N_\text{f})}+\frac{1}{8N_{\text{f}}^2(4\pi)^2}(1-N_\text{f}\lambda_0-\lambda_0),
\end{equation}
with $\lambda_0=\log \left({M_0^2}/{\mu^2}\right)$. 
%As can be seen, this implies that there are $1/N_{f}$ and $1/N_{f}^2$ terms that appear because $M_{eta'}^2 \propto N_{f}$. 
In the large $N_\text{c}$ expansion,  $1/N_\text{f}$ and $1/N_\text{f}^2$ terms remain implicit in the SU($N_\text{f}$) LECs . For this reason, we cannot assume $L_R^{(1)}\Big\rvert_{\text{SU}(N_\text{f})}$ to have a simple $N_\text{f}$ dependence.

\section{Finite-volume formalism} \label{sec:finitevolume}
To compute two-pion scattering amplitudes from lattice results, we use the standard approach:  L\"uscher formalism~\cite{Luscher:1986pf,Luscher:1990ux}. This technique relates the finite-volume spectrum of QCD, obtained from Euclidean correlation functions, to the two-meson scattering amplitude.

We are interested in the simpler formalism for the $s$-wave interaction of two identical (pseudo)scalar particles. The relation reads:
\begin{equation}\label{eq:FullLuscher}
k\cot\delta_0=\frac{1}{\pi L}\mathcal{Z}\left(\frac{Lk}{2\pi}\right),
\end{equation}
where $\delta_0$ is the $s$-wave phase shift of the interaction, $k$ is their momentum in the center-of-mass (CM) frame, and $\mathcal{Z}$ is the Lüscher generalized zeta function~\cite{Luscher:1986pf}. The energy levels on the lattice are related to $k$ as,
\begin{equation}
E_{2}=\sqrt{k^2+M_\pi^2}.
\end{equation}
Near threshold, i.e., $k/M_\pi\ll 1$, one can parametrize the phase shift using the effective range expansion (ERE),
\begin{equation}\label{eq:ERE}
\frac{k}{M_\pi}\cot\delta_0 = \frac{1}{a_0}\left[1-\frac{1}{2}(M_\pi^2a_0r_0)\left(\frac{k^2}{M_\pi^2}\right)+...\right].
\end{equation}

A related result is the so-called threshold expansion. It is a $1/L$ expansion of the energy shift of the ground---or threshold---state of the two particles. This was worked out through $\mathcal{O}(L^{-5})$ in~\cite{Luscher:1986pf}, while the $\mathcal{O}(L^{-6})$ correction was added in Refs.~\cite{Beane:2007qr,Hansen:2015zta,Romero-Lopez:2020rdq}:
\begin{multline}\label{eq:thresholdexp}
\Delta E_{2,\text{thr}} = E_{2,\text{thr}} - 2M_\pi =-\frac{4\pi a_0}{(L)^3}\left[1
+\left(\frac{a_0}{\pi L}\right)\mathcal{I}
+\left(\frac{a_0}{\pi L}\right)^2(\mathcal{I}^2-\mathcal{J})
\right.\\\left.-\left(\frac{a_0}{\pi L}\right)^3(-\mathcal{I}^3+3\mathcal{I}\mathcal{J}-\mathcal{K})
+\frac{2\pi r_0a_0^2}{L^3}
+\frac{\pi a_0}{M_\pi^2L^3}
\right]+\mathcal{O}(L^{-7}).
\end{multline}
Here, $\mathcal{I}, \mathcal{J}$ and $\mathcal{K}$ are known numerical coefficients that can be found, e.g., in Ref.~\cite{Beane:2007qr}. Note that, up to $\mathcal{O}(L^{-5})$, there is a one-to-one relation between the scattering length and the energy shift.

While the full formalism in Eq.~(\ref{eq:FullLuscher}) can always be applied, the threshold expansion should only be used if $a_0/L \ll 1$. In this work we compare both approaches, and check the convergence of  Eq.~(\ref{eq:thresholdexp}). We indeed find examples where the threshold expansion converges poorly.

\section{Lattice setup}\label{sec:setup}
The lattice setup of this work is similar to that of Refs.~\cite{Hernandez:2019qed,Donini:2020qfu}. Our ensembles have been generated with the  HiRep~\cite{DelDebbio:2008zf,DelDebbio:2009fd} code. We use the Iwasaki gauge action~\cite{Iwasaki:1983iya}, and $\mathcal{O}(a)$-improved Wilson fermions in the sea. The value of $c_\text{sw}$ is chosen as described in Ref.~\cite{Hernandez:2019qed}. That is, for $N_\text{c}=3$ we consider the one-loop result boosted by the plaquette, and we keep it constant across other values of $N_\text{c}$.  In all cases, we use periodic boundary conditions. A summary of the simulation parameters is given in Table~\ref{tab:ensembles}.

\begin{table}[b!]
\centering
\begin{tabular}{|c|c|c|c|c|c|c|}
\hline
  Ensemble& $L^3 \times T$ &$\beta$ & $c_\text{sw}$ &$am^\text{s}$ & $am^\text{v}$ & $a \mu_0$  \\ \hline 
3A10 & $20^3 \times 36$ &\multirow{ 5}{*}{1.778}&\multirow{ 5}{*}{1.69}& $-0.4040$ & $-0.4214$  & 0.01107   \\ %\cline{1-2} \cline{5-7} 
3A11 & $24^3 \times 48$ && & $-0.4040$ & $-0.4214$ & 0.01107   \\ %\cline{1-2} \cline{5-7} 
3A20 &$24^3 \times 48$ &&& $-0.4060$   & $-0.4196$ & 0.00781    \\ %\cline{1-2} \cline{5-7} 
3A30 &$24^3 \times 48$ &&& $-0.4070$   & $-0.4187$ & 0.00632   \\ %\cline{1-2}  \cline{5-7} 
3A40 &$32^3 \times 60$ &&& $-0.4080$   & $-0.4163$ & 0.00513  \\ \hline 

3B10 & $24^3 \times 48$ &\multirow{ 2}{*}{1.820} &\multirow{ 2}{*}{1.66}& $-0.3915$  & $-0.4035$ & 0.00825 \\ %\cline{1-2}  \cline{5-7} 
3B20 & $32^3 \times 60$ & & & $-0.3946$ & $-0.4011$ & 0.00431     \\ \hline   

3C10 & $24^3 \times 48$ &\multirow{ 2}{*}{1.850}&\multirow{ 2}{*}{1.64}& $-0.3817$ & $-0.3934$ & 0.00870  \\ %\cline{1-2}  \cline{5-7} 
3C20 & $32^3 \times 60$ & & & $-0.3847$ &$-0.3921$ & 0.00512      \\ \hline   

4A10 &$20^3 \times 36$ &\multirow{ 4}{*}{3.570}&\multirow{ 4}{*}{1.69} &$-0.3725$&  $-0.4163$ & 0.00513 \\ %\cline{1-2}  \cline{5-7} 
4A20 &$24^3 \times 48$& && $-0.3752$ & $-0.3865$& 0.00844 \\ %\cline{1-2}  \cline{5-7} 
4A30 &$24^3 \times 48$& && $-0.3760$ & $-0.3865$& 0.00778 \\ %\cline{1-2} \cline{5-7} 
4A40 &$32^3 \times 60$& && $-0.3780$ & $-0.3851$& 0.00546\\ \hline 
5A10 &$20^3 \times 36$& \multirow{ 4}{*}{5.969}&\multirow{ 4}{*}{1.69} &$-0.3458$ & $-0.3611$ &0.01225 \\ %\cline{1-2}  \cline{5-7} 
5A20 &$24^3 \times 48$& && $-0.3490$ & $-0.3611$& 0.00906   \\ %\cline{1-2} \cline{5-7} 
5A30 &$24^3 \times 48$& && $-0.3500$ & $-0.3607 $& 0.00824  \\ %\cline{1-2} \cline{5-7} 
5A40 &$32^3 \times 60$& && $-0.3530$ & $-0.3596 $& 0.00509   \\ \hline 
6A10 &$20^3 \times 36$ &\multirow{ 4}{*}{8.974}&\multirow{ 4}{*}{1.69} &$ -0.3260$ & $-0.3415$ & 0.01298   \\ %\cline{1-2} \cline{5-7} 
6A20 &$24^3 \times 48$& && $-0.3300$ &$ -0.3414$ & 0.00956 \\ %\cline{1-2}  \cline{5-7} 
6A30 &$24^3 \times 48$& && $-0.3311$  & $-0.3414$ & 0.00803 \\ %\cline{1-2}  \cline{5-7} 
6A40 &$32^3 \times 60$& && $-0.3340$ & $-0.3409$ & 0.00542 \\ \hline 
\end{tabular}
\caption{Summary of ensembles used in this work. $L$ and $T$ indicate the number of points in the spatial and temporal extent, respectively. $\beta$ is the gauge coupling, $c_\text{sw}$ is the $\mathcal{O}(a)$-improvement coefficient, and $am^\text{s}$ is the bare mass of the Dirac operator in the sea sector. Finally, $am^\text{v}$ and $a \mu_0$ are the parameters for the mixed-action setup---the bare mass and the bare twisted mass, in this same order.  }

\label{tab:ensembles}
\end{table}

In this work, we are interested in spectral quantities, so the use of Wilson fermions in the sea is appropriate.  Another option, along the lines of our previous work~\cite{Hernandez:2019qed,Donini:2020qfu}, is to use a mixed-action setup with twisted-mass fermions on the valence. In order to achieve maximal twist, we tune the PCAC valence mass to a value compatible with zero, and we tune the bare twisted mass, $a\mu_0$, so that the valence pion mass, $M_\pi^\text{v}$, matches the sea value, $M_\pi^\text{s}$. There are various advantages for the latter choice. First, automatic $\mathcal{O}(a)$ improvement is achieved.\footnote{Up to residual $\mathcal{O}(a \alpha_\text{s}^2)$ effects coming from the sea quark mass~\cite{Bussone:2019mlt,Ugarrio:2018ghf} and the one-loop value for $c_\text{sw}$.} Second, the pion decay constant can be estimated from the matrix elements of the pseudoscalar current, $P$, without the need of renormalization constants:
\begin{equation}\label{eq:decayconstant}
F_\pi=\frac{\sqrt{2}\mu_0\langle0|P|\pi\rangle _{\text{bare}}}{M_\pi^2},
\end{equation}
which is relevant for the chiral fits of this work. The combination of both lattice regularizations will be useful to quantify discretization effects.

In Table~\ref{tab:massdecayxi}, we summarize our results for the single-pion quantities. We introduce a new quantity, the chiral paremter, $\xi$, defined as
\begin{equation}\label{eq:chiralparameter}
\xi=\frac{M_\pi^2}{(4\pi F_\pi)^2},
\end{equation}
which will be useful when matching our results to ChPT. Note that the leading finite-volume effects from Refs.~\cite{Gasser:1987ah, Colangelo:2005gd} have been taken into account. The typical size of these corrections is found to be of $\sim1\%$. In addition, the errors in the tuning of the PCAC mass in the mixed-action setup, measured by $|m_\text{PCAC}^\text{v}/\mu_0|$, are at most at the few-percent level. This implies that the effects in our results are well below the statistical precision.

\begin{table}[h!]
\centering
\setlength{\tabcolsep}{4.5pt}
\begin{tabular}{|c|c|c|c|c|c|c|}
\hline
Ensemble& $aM_{\pi,L}^\text{s}$ & $aM_{\pi,L}^\text{v}$ & $|m_\text{PCAC}^\text{v}/\mu_0|$ & $aF_{\pi,L}$ & $\xi$ & $\xi_L$ 	\\ \hline
3A10 & 0.222(3) 		& 0.2211(23)	 &  $0.007$(11) 	& 0.0449(4) 	 	& 0.159(6)     	& 0.154(6)	\\
3A11 & 0.2150(15) 	& 0.2185(10)	 &  0.002(12) 		& 0.0452(3) 	 	& 0.150(3)     	& 0.148(3)	\\
3A20 & 0.1853(13) 	& 0.1830(8)	 &  $0.029$(16) 	& 0.0409(3) 	 	& 0.130(3)     	& 0.1267(25)	\\
3A30 & 0.1611(16) 	& 0.1616(8)	 &  0.003(16) 		& 0.0378(3) 	& 0.123(3)     	& 0.1157(25)	\\
3A40 & 0.1419(10) 	& 0.1423(6)	 &  0.002(12) 		& 0.03577(14) 	& 0.1021(14)   	& 0.1003(14)	\\  \hline
3B10 & 0.1751(11) 	& 0.1764(9)	 &  0.005(8) 		& 0.03626(23) 	& 0.157(3)     	& 0.150(3)	\\
3B20 & 0.1189(8) 	& 0.1221(6)	 &  0.001(7) 		& 0.03121(13) 	& 0.1015(16)   	& 0.0969(16)	\\  \hline
3C10 & 0.1756(18) 	& 0.1759(18)	 &  0.007(13) 		& 0.0336(4) 	 	& 0.183(7)     	& 0.174(7)	\\
3C20 & 0.1308(13) 	& 0.1289(12)	 &  0.007(19) 		& 0.02866(24) 	& 0.134(4)    	& 0.128(4)	\\  \hline
4A10 & 0.2044(13) 	& 0.2035(15)	 &  0.001(10) 		& 0.0521(4) 	 	& 0.100(3)   	& 0.0968(25)\\
4A20 & 0.1805(8) 	& 0.1799(6) 	 &  0.002(8) 		& 0.05103(16) 	& 0.0803(9)   	& 0.0787(9)	\\
4A30 & 0.1707(7) 	& 0.1730(5)	 &  0.005(8) 		& 0.04952(19) 	& 0.0793(9)   	& 0.0773(9)	\\
4A40 & 0.1399(8) 	& 0.1419(8) 	 &  0.002(11) 		& 0.0464(3) 	 	& 0.0599(12)   	& 0.0593(12)	\\  \hline
5A10 & 0.2125(11) 	& 0.2126(8)	 &  0.014(5) 		& 0.06154(22) 	& 0.0772(10)   	& 0.0756(10)	\\
5A20 & 0.1803(6) 	& 0.1798(5)	 &  0.005(11) 		& 0.05846(24) 	& 0.0609(6)    	& 0.0599(6)	\\
5A30 & 0.1707(6) 	& 0.1715(6)	 &  0.009(9) 		& 0.0570(3) 	 	& 0.0584(9)   	& 0.0573(9)	\\
5A40 & 0.1331(5) 	& 0.1330(4)	 &  0.007(8) 		& 0.05306(15) 	& 0.0403(3)    	& 0.0398(3)	\\  \hline
6A10 & 0.2147(7) 	& 0.2142(6)	 &  $0.006$(4) 	& 0.06874(21) 	& 0.0625(6)    	& 0.0615(6)	\\
6A20 & 0.1798(6) 	& 0.1802(4)	 &  0.003(5) 		& 0.06582(21) 	& 0.0481(4) 		& 0.0475(4)	\\
6A30 & 0.1685(7) 	& 0.1666(5)	 &  0.010(11) 		& 0.06324(23) 	& 0.0447(5)	   	& 0.0439(5)	\\
6A40 & 0.1353(3)		& 0.1347(5)	 &  0.001(7) 		& 0.05950(13)	& 0.0327(3)	   	& 0.0324(3)	\\  \hline
\end{tabular}
\caption{ Results for the pion mass for the sea and the valence sector ($M_\pi^{\text{s}}$ and $M_\pi^\text{v}$, respectively), the pion decay constant, the PCAC mass in units of the twisted mass, and the chiral parameter, as defined in Eq.~(\ref{eq:chiralparameter}). The ``$L$'' subscript indicates that the leading finite-volume effects 
estimated from ChPT
have been taken into account~\cite{Gasser:1987ah, Colangelo:2005gd}.}
\label{tab:massdecayxi}
\end{table}

\subsection{Scale setting}

In order to study the lattice artifacts of the quantities we explore here, it is useful to know the lattice spacing in physical units. For this, we follow the procedure outlined in Ref.~\cite{Hernandez:2019qed}. This requires measuring $t_0/a^2$, for which an improved estimator is computed, $t_0^\text{imp}/a^2$, and $M_\pi^\text{s}$ on each ensemble. Our scale setting condition is then
\begin{equation}
\left( M_\pi^\text{s} \sqrt{t_0} \right)\big \rvert_{M_\text{ref}^\text{s}=420 \text{ MeV}} = 0.309(8),
\end{equation}
where $M_\text{ref}^\text{s}$ is the reference mass. We use the physical value:
\begin{equation}
\sqrt{t_0}\big \rvert^{N_\text{f}=4}_{M_\text{ref}^\text{s}} = 0.145(4), \label{eq:t0Mref}
\end{equation}
obtained from extrapolating existing lattice results~\cite{Bruno:2013gha,Sommer:2014mea,Bruno:2016plf} at $N_\text{f}=2,3$ to $N_\text{f}=4$. Indeed, the uncertainty in Eq.~(\ref{eq:t0Mref}) will turn out to be the largest in the determination of the lattice spacing.

For the ``A'' ensembles detailed in Table~\ref{tab:ensembles} the scale setting was carried out in Ref.~\cite{Hernandez:2019qed}, and we do not repeat it here. The result was $a = 0.075(2)$ fm for $N_\text{c}=3$, and very close for $N_\text{c}>3$. Here we focus on the newer and finer ``B'' and  ``C'' ensembles. The necessary values of $t_0^\text{imp}/a^2$ are given in Table~\ref{tab:t0}. Using the ChPT expectation for the flow scale~\cite{Bar:2013ora}, we find for each set of ensembles the value of $t_0^\text{imp}/a^2$ at the reference mass, $M_\text{ref}$. The final results for the lattice spacing can be obtained by combining the value of $(t_0^\text{imp}/a^2) \rvert_{M_\text{ref}}$ and Eq.~(\ref{eq:t0Mref}). From the last column of Table~\ref{tab:t0}, and assuming $\mathcal{O}(a^2)$ scaling, the expected reduction of discretization effects from the ensembles ``A'' to ``C'' is of about $40\%$.\vspace{-.2 cm}

\begin{table}[h!]
\centering
\vspace{-0.15 cm}
%\fontsize{11pt}{11pt}\selectfont
\begin{tabular}{|c|c|c|c|}
\hline
Ensemble& $t_0^\text{imp}/a^2$	& $(t_0^\text{imp}/a^2) \rvert_{M_\text{ref}}$ & $a$/fm\\ \hline
3B10 & 4.61(10)&\multirow{2}{*}{4.97(6)} & \multirow{2}{*}{0.065(2)}\\
3B20 & 5.13(3)& & \\  \hline
3C10 & 5.534(12)&\multirow{2}{*}{6.10(15)} & \multirow{2}{*}{0.059(2)}	\\
3C20 & 6.05(14)& &	\\  \hline % \hline
\end{tabular}
\caption{ Summary of the quantities entering in the scale setting of the ``B'' and  ``C'' ensembles, which are the new ones generated for this work. Following the procedure of Ref.~\cite{Hernandez:2019qed}, we use the clover definition of the energy density of the field strength tensor and apply the improvement of Ref.~\cite{Fodor:2014cpa}. In addition, we interpolate to the reference mass, $M_\text{ref}=420$ MeV, and calculate the value of the lattice spacing in physical units using Eq.~(\ref{eq:t0Mref}). }
\label{tab:t0}
\end{table}

\subsection{Extraction of energy levels}

In order to determine the energy levels on the lattice, we need to measure the appropriate two-pion correlation functions.\vspace{-0.05 cm}
\begin{equation}
C_R(t) = \langle O^{\dagger}_R(t) O_R(0) \rangle,\vspace{-0.05 cm}
\end{equation}
where  the two-pion operators for the channels of interest are:\vspace{-0.05 cm}
\begin{equation}
 O_{SS}(t) = \pi^+(t) \pi^+(t), \quad\quad O_{AA}(t) = \frac{1}{\sqrt{2}}\left[ \pi^+(t) D_s^+(t) - K^+(t) D^+(t) \right].\vspace{-0.05 cm}
\end{equation}
Here, the single-meson interpolators are projected to zero momentum, that is,
\begin{equation}
\pi^+(t) = - \sum_{\bm{ x}}  \bar d({\bm{ x}},t) \gamma_5 u ({\bm{ x}},t),\vspace{-0.05 cm}
\end{equation}
and analogously for $K^+, D^+ $ and $D_s^+$, with the corresponding quark flavors. As we have seen in Sec.~\ref{sec:ChPT} the two possible Wick contractions are those in
Fig.~\ref{fig:contractions} and Eq.~(\ref{eq:wickcontractions}).

In the limit of infinite time extent, the two-pion correlation functions must decay as exponentials in Euclidean time. However, the propagation of particles around the periodic boundary in the time direction gives rise to additional contributions to the correlator. These are the so-called thermal effects, which can be sizeable around $t \simeq T/2$. For the zero-momentum correlators, the leading thermal pollution term is time-independent 
\begin{equation}
C_R(t) = A \cosh \left[ E_R (t-T/2) \right] + B,\vspace{-0.05 cm}
\end{equation}
where $A$ and $B$ are real and positive constants related to the matrix elements of the operator, and $E_R$ is the two-pion energy in channel $R$. A useful observable is the following ratio~\cite{PhysRevD.75.094502}:\vspace{-0.05 cm}
\begin{equation}\label{eq:ratio}
R(t)=\frac{C_R(t+1)-C_R(t-1)}{C_\pi^2(t+1)-C_\pi^2(t-1)},
\end{equation}
which eliminates the constant term, and empirically seems to cancel contributions from excited states. The asymptotic time dependence of this ratio is
\begin{equation}
R(t)=R_0\left[\cosh(\mathrm{\Delta} E_R t')+\sinh(\Delta E_R t')\coth(2M_\pi t')\right], \label{eq:ratioeq}
\end{equation}
being $t'=t-T/2$ and $\mathrm{\Delta} E_R=E_R-2M_\pi$, with $R_0$ an arbitrary normalization.   

\begin{figure}[h!]
\centering
\subfigure%
{\includegraphics[width=0.48\textwidth,clip]{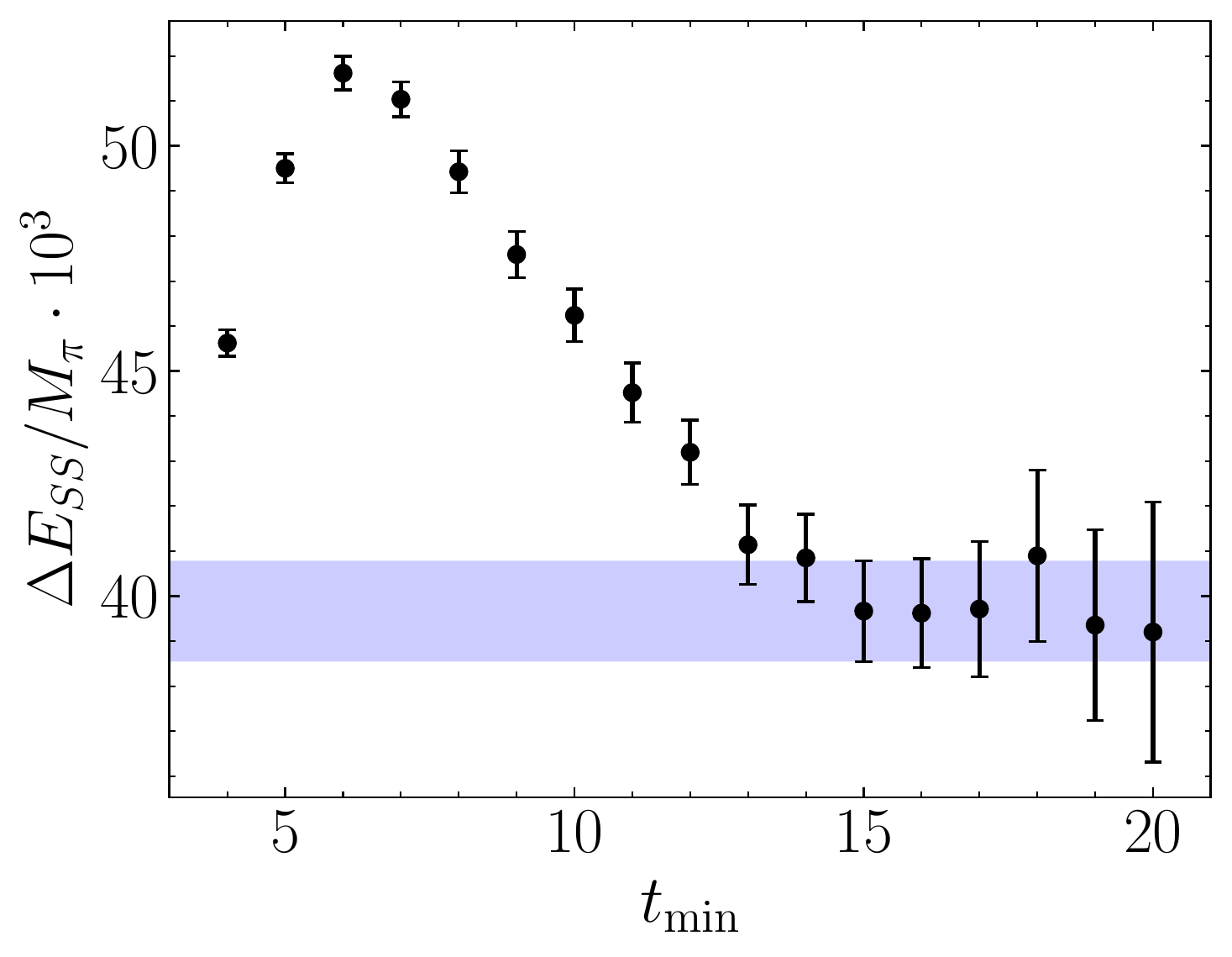}}\hfill
\subfigure%
{\includegraphics[width=0.519\textwidth,clip]{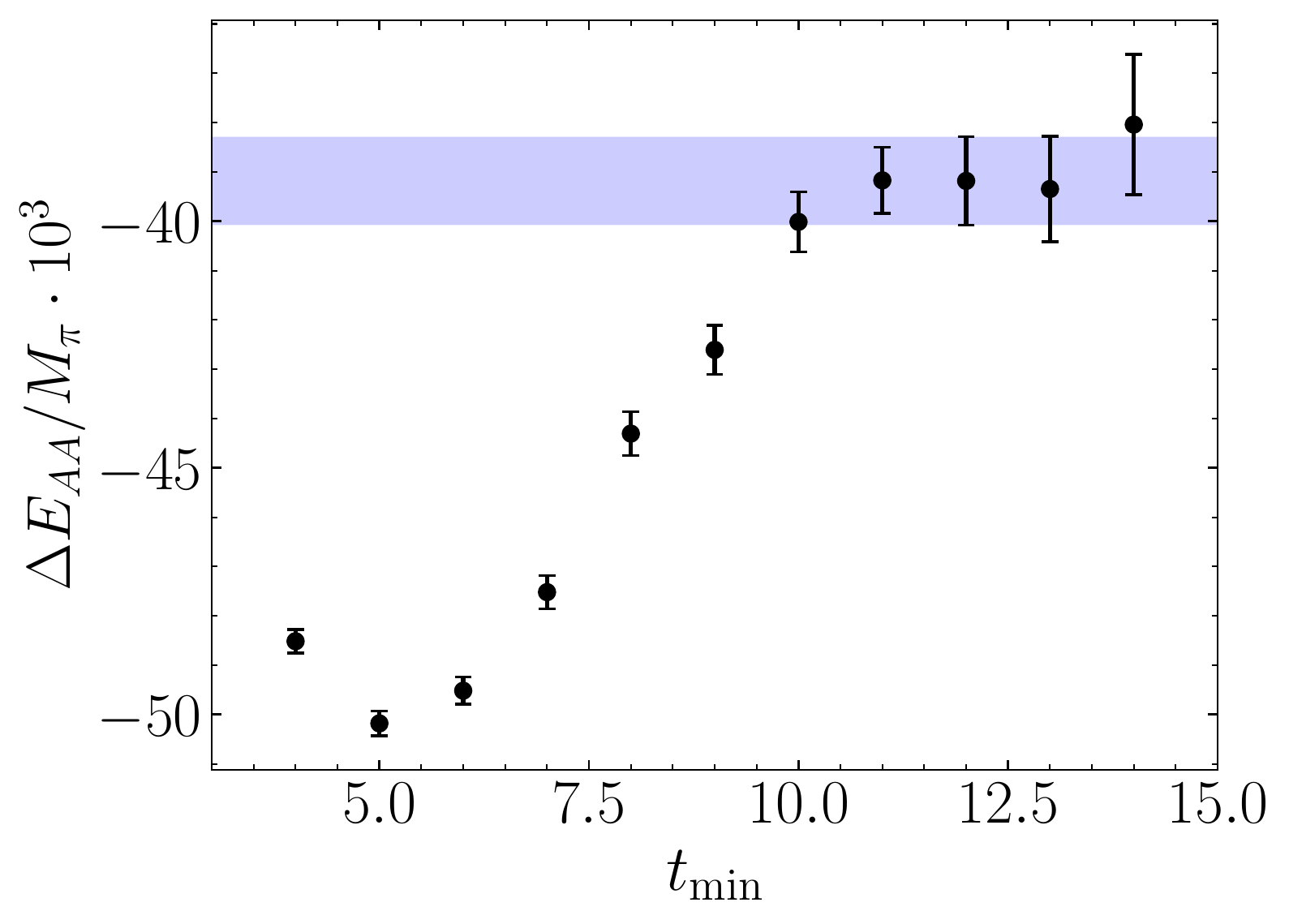}}
\caption{Examples of the energy shift determination for the two channels and different ensembles:  4A20 for the $SS$ channel (right) and 6A10 for the $AA$ channel (right). }
\label{fig:plateaux}
\end{figure}

We perform correlated fits to Eq.~(\ref{eq:ratioeq}), and estimate the errors using bootstrapping. In these fits, we fix $M_\pi$ to its value obtained from the one-pion correlator in the corresponding bootstrap sample. In addition, we take into account autocorrelations in the Markov chain by using block lengths that as several times larger than the integrated autocorrelation time. We also consider correlations between time slices using the covariance matrix computed from all initial blocks. The fit range is taken to be $[t_\text{min}, T/2-1]$, where $t_\text{min}$ is chosen such that the resulting value of $\Delta E_R$ lies on a plateau. The results for the energies are provided in the second column of Table~\ref{tab:energies}. In addition, two examples of these plateaux are shown in Fig.~\ref{fig:plateaux}---one corresponding to each channel.

\begin{table}[tph!]
\centering

\begin{tabular}{|c|c|c|c|c|}
\hline
Ensemble& $\Delta E_{SS}/M_\pi$ & $M_\pi a_0^{SS}$ & ${\Delta} E_{AA}/M_\pi$ & $M_\pi a_0^{AA}$ \\ \hline   
3A10 & 0.100(4) 		& -0.494(17)	 &  -0.075(4) 	 & 0.73(4)     \\ 
3A11 & 0.0495(17) 	& -0.442(13)	 &  -0.0458(13) 	 & 0.72(3)     \\ 
3A20 & 0.082(4) 		& -0.416(14)	 &  -0.063(3) 	 & 0.57(4)     \\  
3A30 & 0.114(4) 		& -0.392(11)	 &  -0.082(7) 	 & 0.52(5)     \\   
3A40 & 0.0532(20) 	& -0.324(11)	 &  -0.050(3) 	 & 0.49(3)     \\  \hline  
3B10 & 0.104(3) 		& -0.456(9)	 &  -0.093(3) 	 & 0.81(3)     \\   
3B20 & 0.087(3) 		& -0.324(10)	 &  -0.064(3) 	 & 0.390(24)   \\  \hline  
3C10 & 0.120(6) 		& -0.503(19)	 &  -0.117(5) 	 & 1.02(4)     \\   
3C20 & 0.111(9) 		& -0.45(3) 	 &  -0.082(4) 	 & 0.64(4)     \\  \hline  
4A10 & 0.0825(25) 	& -0.345(9) 	 &  -0.063(3) 	 & 0.445(25)   \\ 
4A20 & 0.0473(14) 	& -0.255(6) 	 &  -0.0410(15) 	 & 0.319(15)   \\  
4A30 & 0.0529(14) 	& -0.252(6)	 &  -0.0456(23) 	 & 0.317(20)   \\   
4A40 & 0.0269(13) 	& -0.179(9) 	 &  -0.0276(11) 	 & 0.236(11)   \\  \hline 
5A10 & 0.0507(11) 	& -0.259(4)	 &  -0.0469(10) 	 & 0.356(10)   \\ 
5A20 & 0.0342(11) 	& -0.192(6)	 &  -0.0334(7) 	 & 0.250(6)    \\  
5A30 & 0.0381(12) 	& -0.186(6)	 &  -0.0323(15) 	 & 0.206(11)   \\   
5A40 & 0.0235(10) 	& -0.132(6)	 &  -0.0209(13) 	 & 0.141(9)    \\  \hline 
6A10 & 0.0402(15) 	& -0.217(7)	 &  -0.0393(9) 	 & 0.295(8)    \\ 
6A20 & 0.0290(13) 	& -0.167(6)	 &  -0.0261(6) 	 & 0.189(5) 	   \\  
6A30 & 0.0301(16) 	& -0.139(7)	 &  -0.0308(16) 	 & 0.177(11)	   \\   
6A40 & 0.0171(5)		& -0.102(3)	 &  -0.0172(10)	 & 0.118(7)	   \\  \hline  

\end{tabular}
\caption{  Results for the energy shifts $\mathrm{\Delta} E_R = E_R - 2M_\pi$ for the two channels, and the corresponding scattering lengths computed using Eq.~(\ref{eq:thresholdexp}) to $\mathcal{O}(L^{-5})$. }
\label{tab:energies}
\end{table}

\section{Results for $\pi\pi$ scattering}\label{sec:scattlen}

In this section, we use our results for ground state energy levels, together with the $1/L$ expansion of the ground state---given in Eq.~(\ref{eq:thresholdexp})---to obtain the $s$-wave scattering length. Since this is a perturbative expansion valid when $a_0/L \ll 1$, we test whether this condition is satisfied a posteriori. We also study discretization effects in the scattering amplitude.

We start by computing the scattering length to $\mathcal{O}(L^{-5})$. The results are summarized in the third and fifth columns of Table~\ref{tab:energies}. As can be seen, the $SS$ channel is repulsive, while the $AA$ one is attractive. Qualitatively, the scattering lengths are of the same order for both channels, but with opposite signs. This is expected from ChPT and large $N_\text{c}$, as discussed in Sec.~\ref{sec:ChPT}.

\begin{figure}[h!]
\centering
\subfigure%
[$SS$ channel]{\includegraphics[width=0.5115\textwidth,clip]{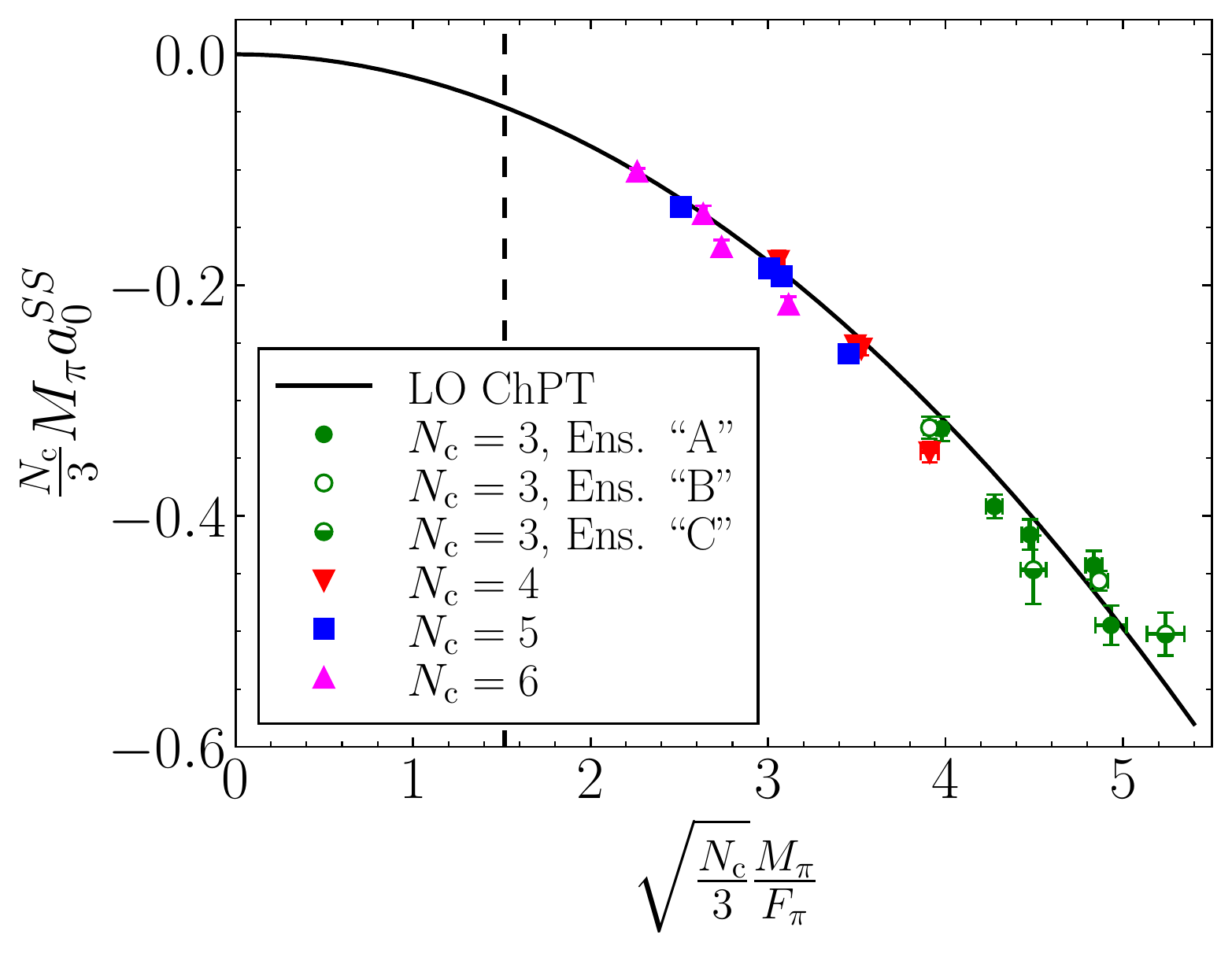}}\hfill
\subfigure%
[$AA$ channel]{\includegraphics[width=0.4885\textwidth,clip]{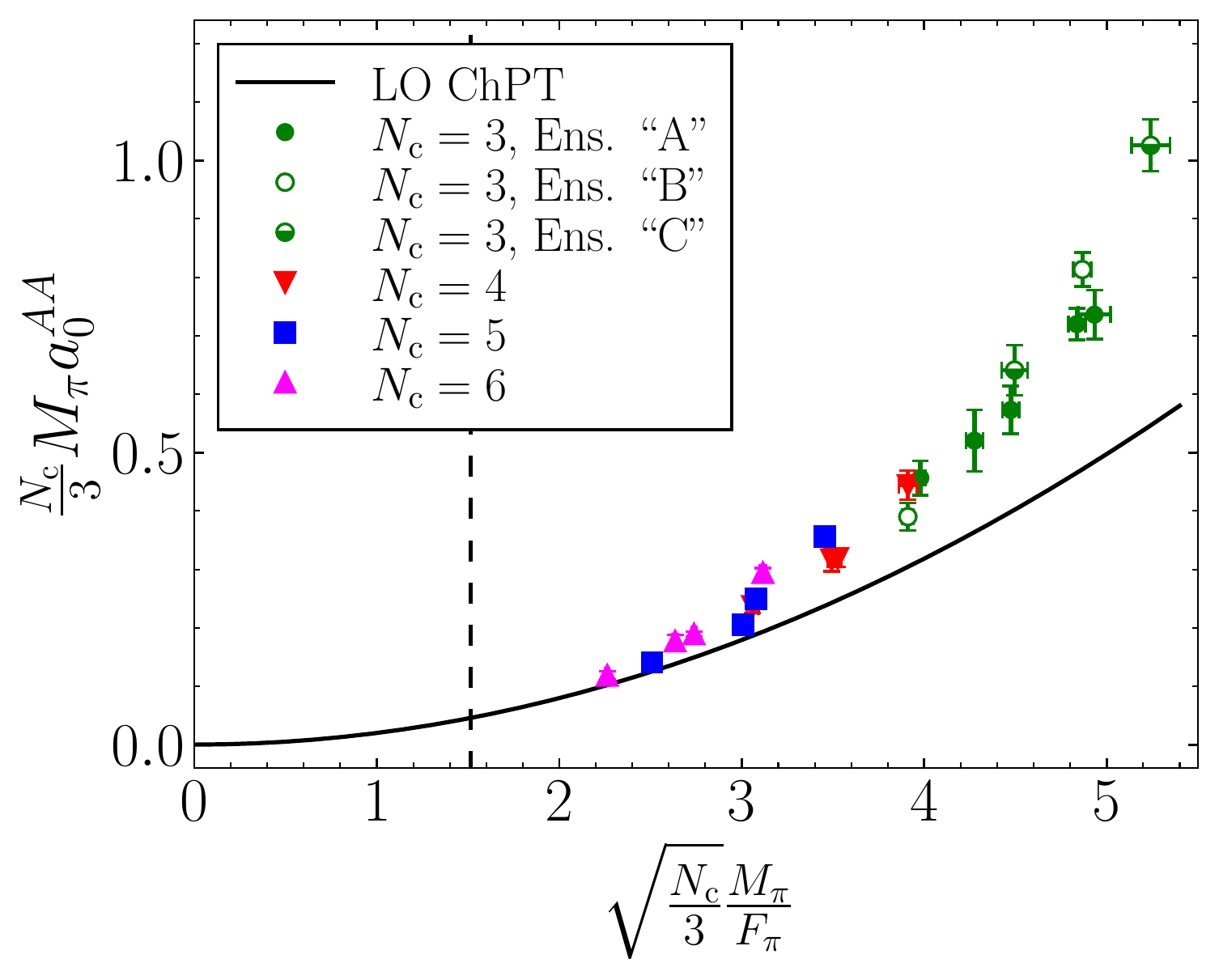}}
\caption{Results for the $s$-wave scattering length obtained using the threshold expansion to $\mathcal{O}(L^{-5})$, against the LO predictions from ChPT (black solid line). Both axis have been multiplied by a $N_\text{c}$-dependent factor to eliminate leading $N_\text{c}$ dependencies. The physical point is indicated by a vertical dashed line. }
\label{fig:LOscatteringlength}
\end{figure}

In Fig.~\ref{fig:LOscatteringlength} we  compare the scattering length against the LO prediction of ChPT. We observe good agreement for the $SS$ channel, suggesting that higher order corrections are small in this channel. The situation is different for the $AA$ channel, where deviations from the LO are sizeable. We now study a possible lack of convergence of the threshold expansion.

%{\bf TO DO: discuss why convergence may be an issue and we will test it in the next section.}

\subsection{Convergence of threshold expansion} \label{sec:convergence}

As explained in Sec.~\ref{sec:finitevolume}, Eq.~(\ref{eq:thresholdexp}) is an expansion in powers of $a_0/L$. The truncation of the expression to $\mathcal{O}(L^{-5})$ introduces some systematic error, that may not be neglected if $a_0/L \sim 1$. In order to study this effect, a comparison is done between the $\mathcal{O}(L^{-5})$ and the $\mathcal{O}(L^{-6})$ expansions. Since $r_0$ enters at $\mathcal{O}(L^{-6})$, we will estimate its effect based on its LO ChPT prediction given in Eq.~(\ref{eq:EffectiveRangeLO}). Using the value of $\Delta E$ for each ensemble, we collate the results for $M_\pi a_0$ at both orders of the threshold expansion. 

\begin{figure}[h!]
\centering
\subfigure%
[$SS$ channel]{\includegraphics[width=0.49\textwidth,clip]{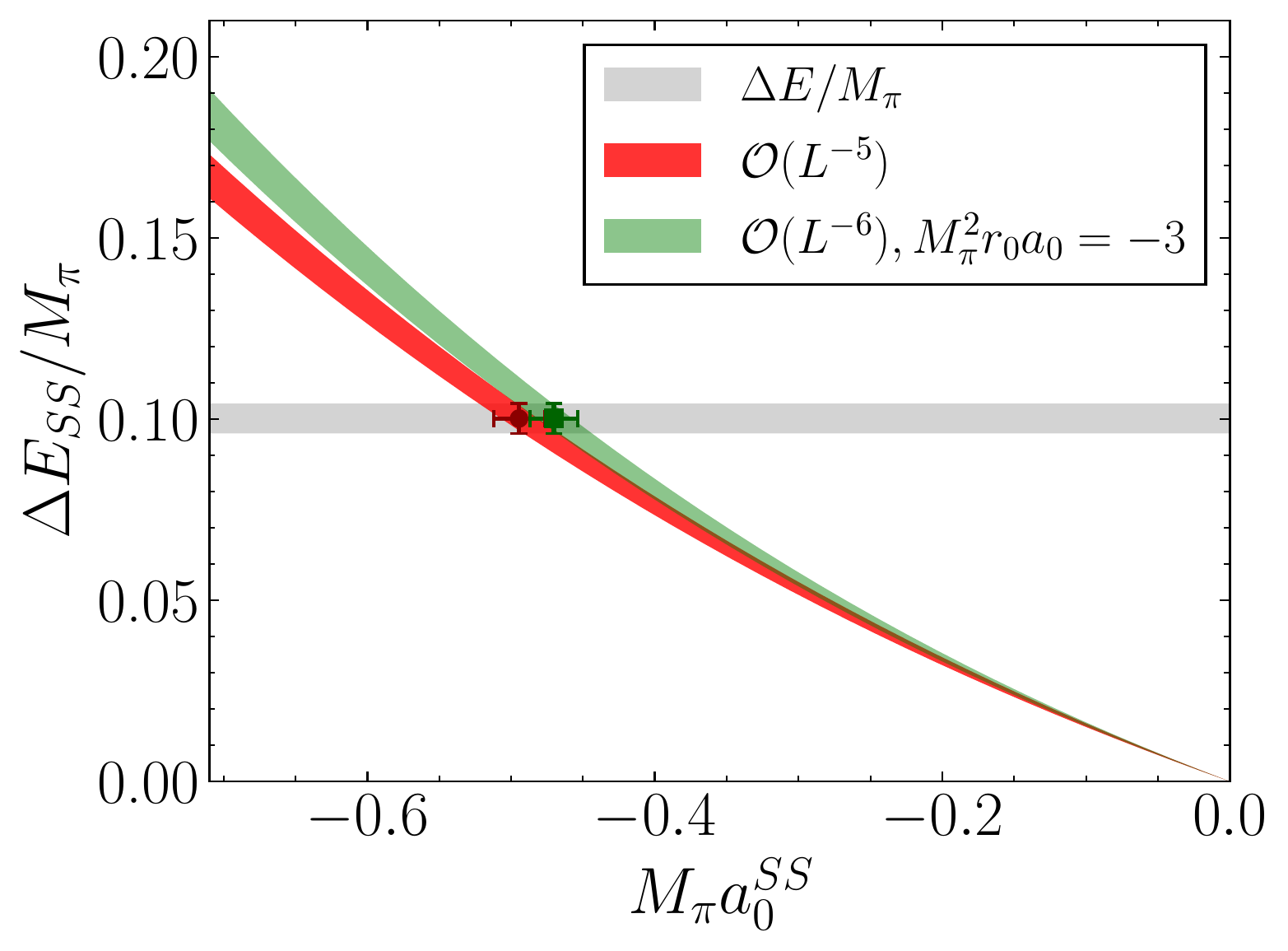}}\hfill
\subfigure%
[$AA$ channel]{\includegraphics[width=0.508\textwidth,clip]{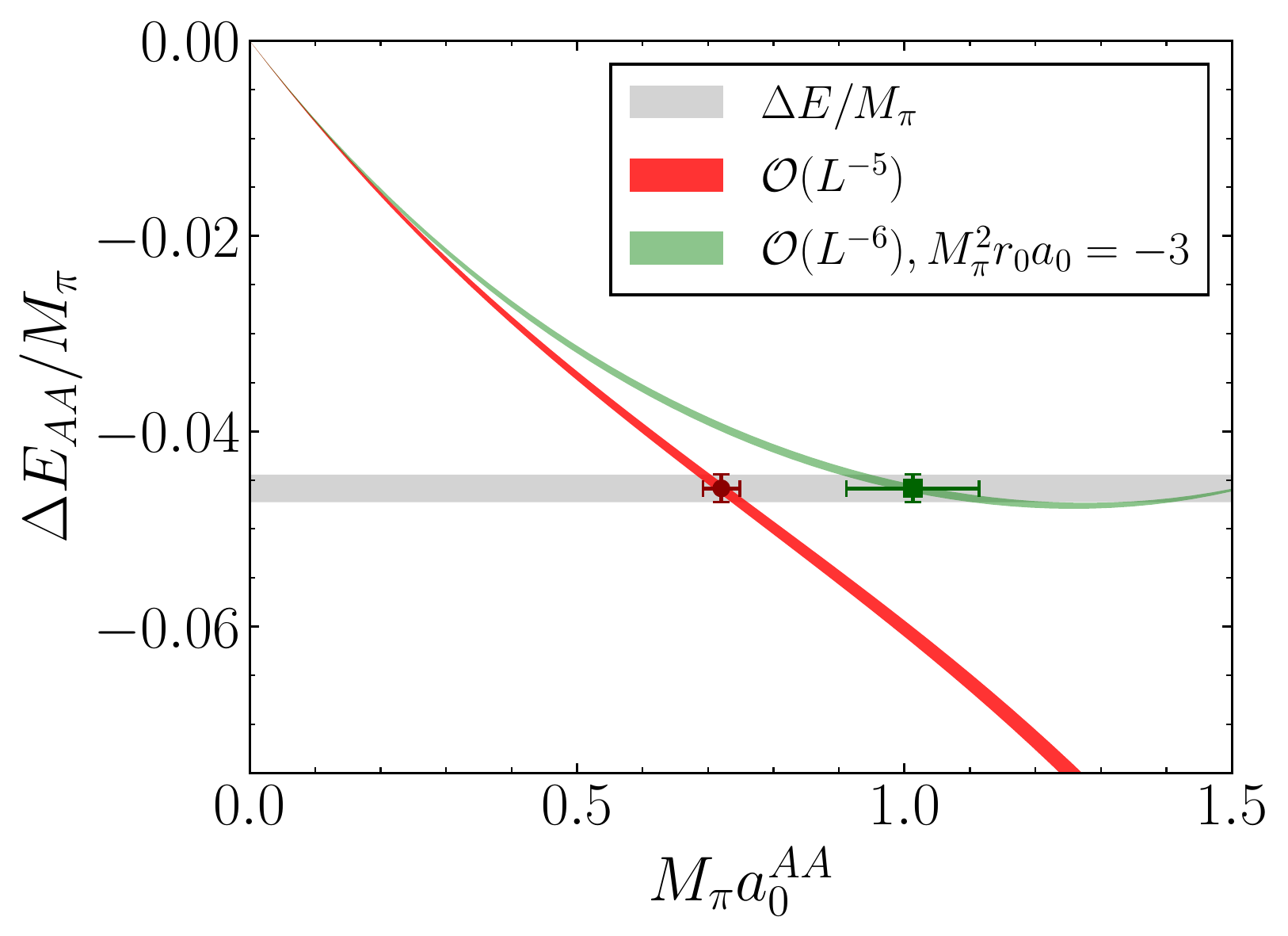}}
\caption{ Comparison between the  $\mathcal{O}(L^{-5})$ (red) and $\mathcal{O}(L^{-6})$ (green) threshold expansion given in Eq.~(\ref{eq:thresholdexp}). The left panel corresponds to the 3A10 ensemble in the $SS$ channel, whereas the right one is the 3A11 ensemble and $AA$ channel. The gray horizontal band is the energy level extracted from the lattice, and the points are the determinations of the scattering length.}
\label{fig:convergence}
\end{figure}

In the case of the $SS$ channel, we observe that for all the ensembles, both assumptions produce very similar results, with discrepancies smaller than 1$\sigma$. On the left panel of Fig.~\ref{fig:convergence} we show this comparison for the 3A10 ensemble. The gray horizontal band corresponds to the lattice $\Delta E_{SS}$ result, while the red and green ones are the $\mathcal{O}(L^{-5})$ and  $\mathcal{O}(L^{-6})$ threshold expansions, respectively, whose width comes from the errors of $M_\pi L$. The red and green points are the results for the scattering length, which are compatible.
\vspace{0.3cm}

\vspace{0.3cm}

The case of the $AA$ channel is more complicated. For the ensembles with smaller $\xi$ and larger $L$, we find acceptable convergence, and compatible results for $a_0^{AA}$ at both orders. However, as $\xi$ is increased we arrive to a point at which convergence fails. An example of this is shown in the right panel of Fig.~\ref{fig:convergence}, with the same color code as before. We conclude that some of our ensembles lie in a regime in which the $1/L$ expansion is no longer valid, and so any results for the scattering lengths may have uncontrolled systematic errors.
In view of this caveat, we avoid the expansion and use instead the full Lüscher formalism. 
%With this, we are able to obtain the energy shift of the two pions directly from lattice results, and match it to ChPT to constrain the $N_\text{c}$ scaling of the relevant LECs.

\subsection{Discretization effects} \label{sec:cutofferrors}

Discretization uncertainties are always a concern in lattice calculations. Previous work in the isospin-2 channel with twisted-mass fermions has shown that discretization errors are not very significant for this channel~\cite{ETM:2015bzg}. However, hints from the two-baryon sector~\cite{Green:2021qol} indicate that this may not always be the same in all scattering observables. Indeed, as we will see, we find significant discretization effects in the $AA$ channel. This is, to our knowledge, the first time that such large effects are seen for meson-meson scattering.

\begin{figure}[h!]
\centering
\subfigure%
[$SS$ channel]{\includegraphics[width=0.49\textwidth,clip]{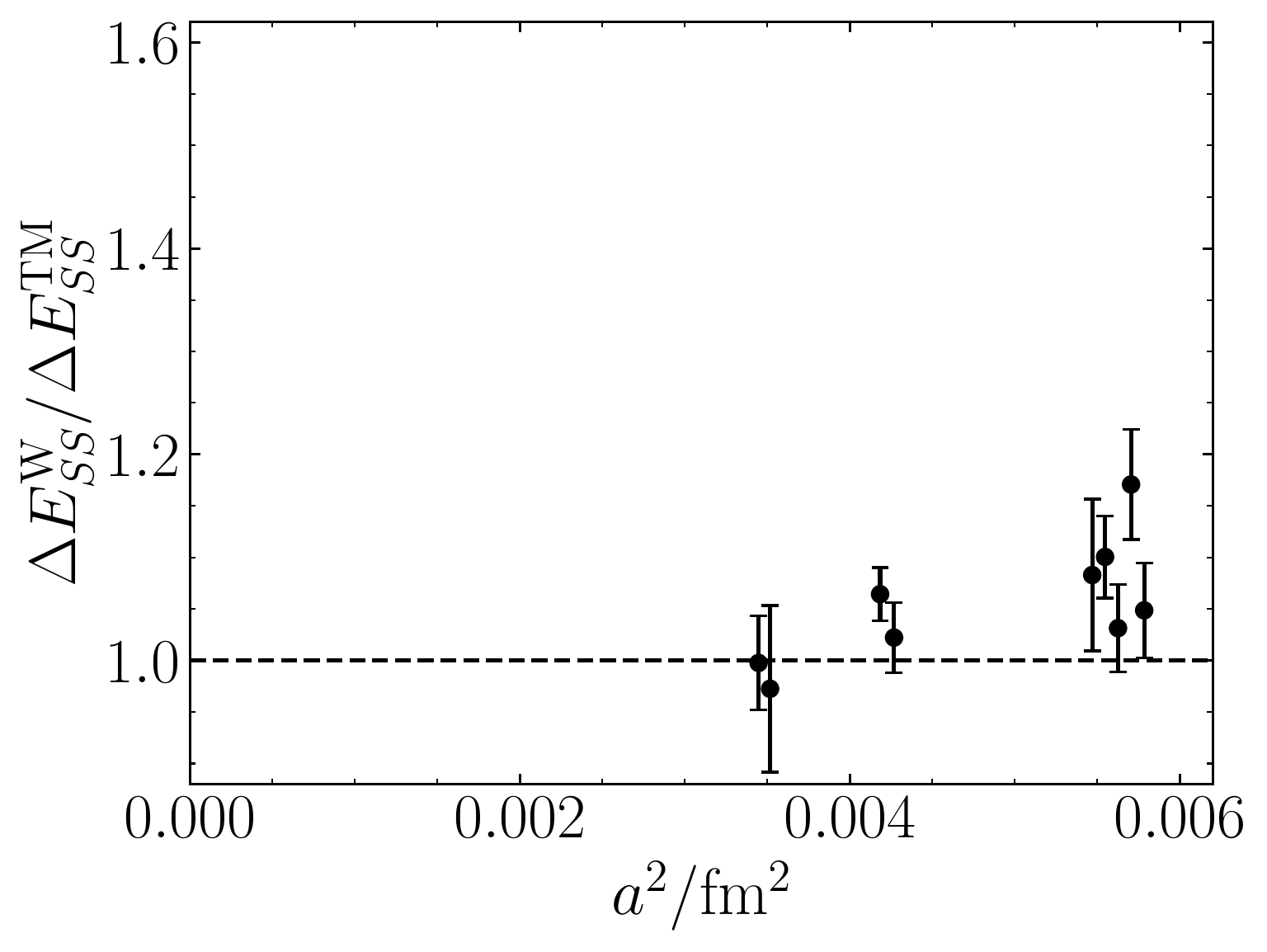}}\hfill
\subfigure%
[$AA$ channel]{\includegraphics[width=0.49\textwidth,clip]{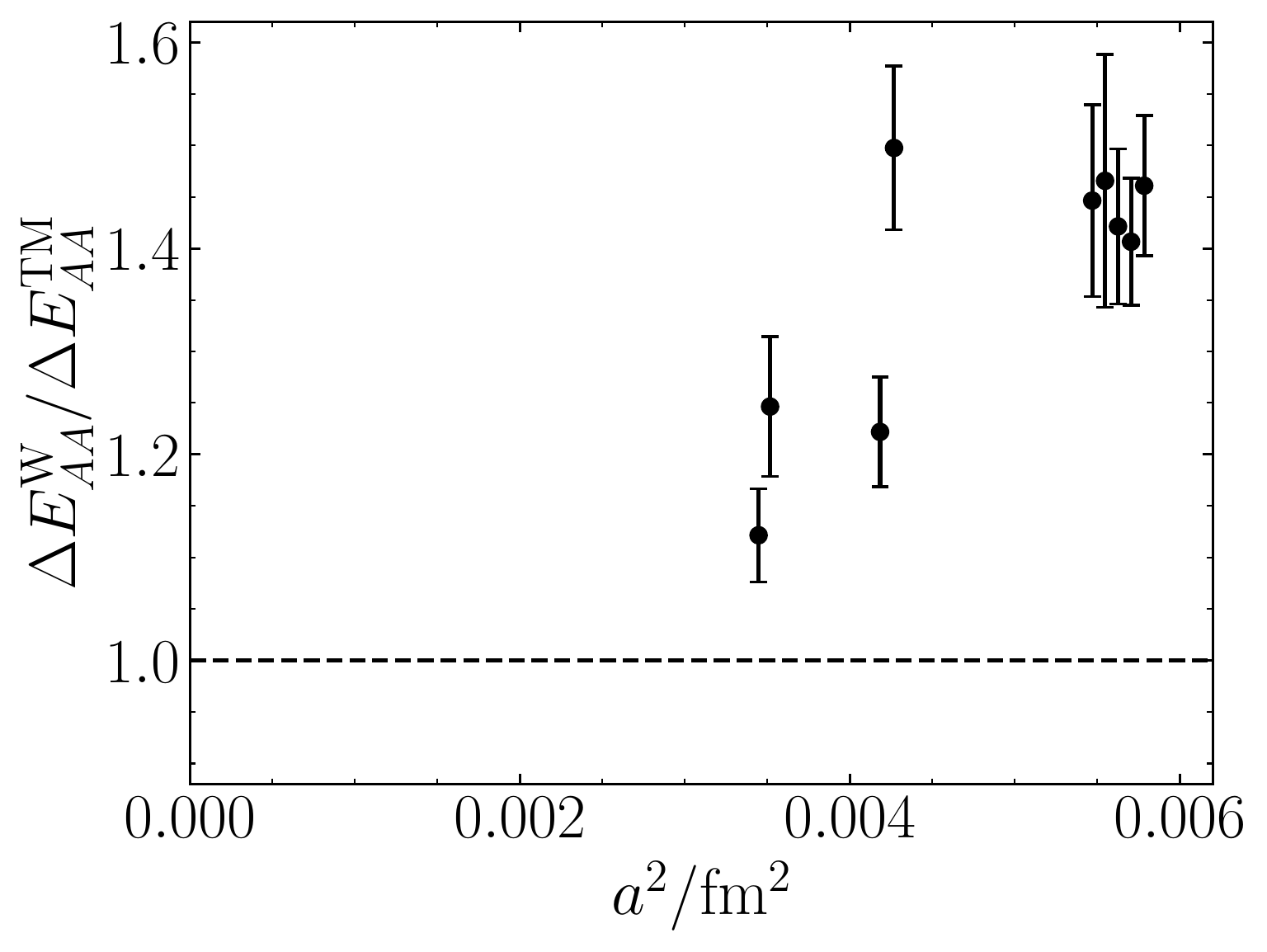}}
\caption{ Dependence on the lattice spacing of the quotient between the energy shifts computed using two different fermion regularizations: the Wilson setup and the mixed-action one. The comparison is done both for the $SS$ (left) and $AA$ (right) channels.}
\label{fig:EnergyComparison}
\end{figure}

We will assess the magnitude of the cutoff effects at $N_\text{c}=3$, since various ensembles at three different lattice spacings are available. First, we compare the results using the pure Wilson and mixed-action setups, which should coincide in the continuum limit. In Fig.~\ref{fig:EnergyComparison} we represent the ratio between the energy shifts determined for both regularizations. We observe that for the $SS$ channel all results are close to unity and thus discretization effects are small. However, this is not the case for the $AA$ channel. We observe large discretization effects, which nevertheless seem to decrease as we approach the continuum limit. Since only the $AA$ energy shifts show relevant cutoff effects, we will focus on this channel henceforth.

To understand the absolute impact of discretization effects in both regularizations  we consider the continuum extrapolation of the scattering phase shift, $k\cot\delta_0^{AA}$. This implies additional difficulties, since the scattering amplitude is a function of both $\xi$ and the CM momentum $k$, which differ for each ensemble. To keep a line of constant physics, we first need to extrapolate/interpolate to a reference point ($k_\text{ref}^2,\xi_\text{ref} $) for each lattice spacing and fermion regularization.

\begin{figure}[h!]
\centering
\subfigure%
{\includegraphics[width=0.7\textwidth,clip]{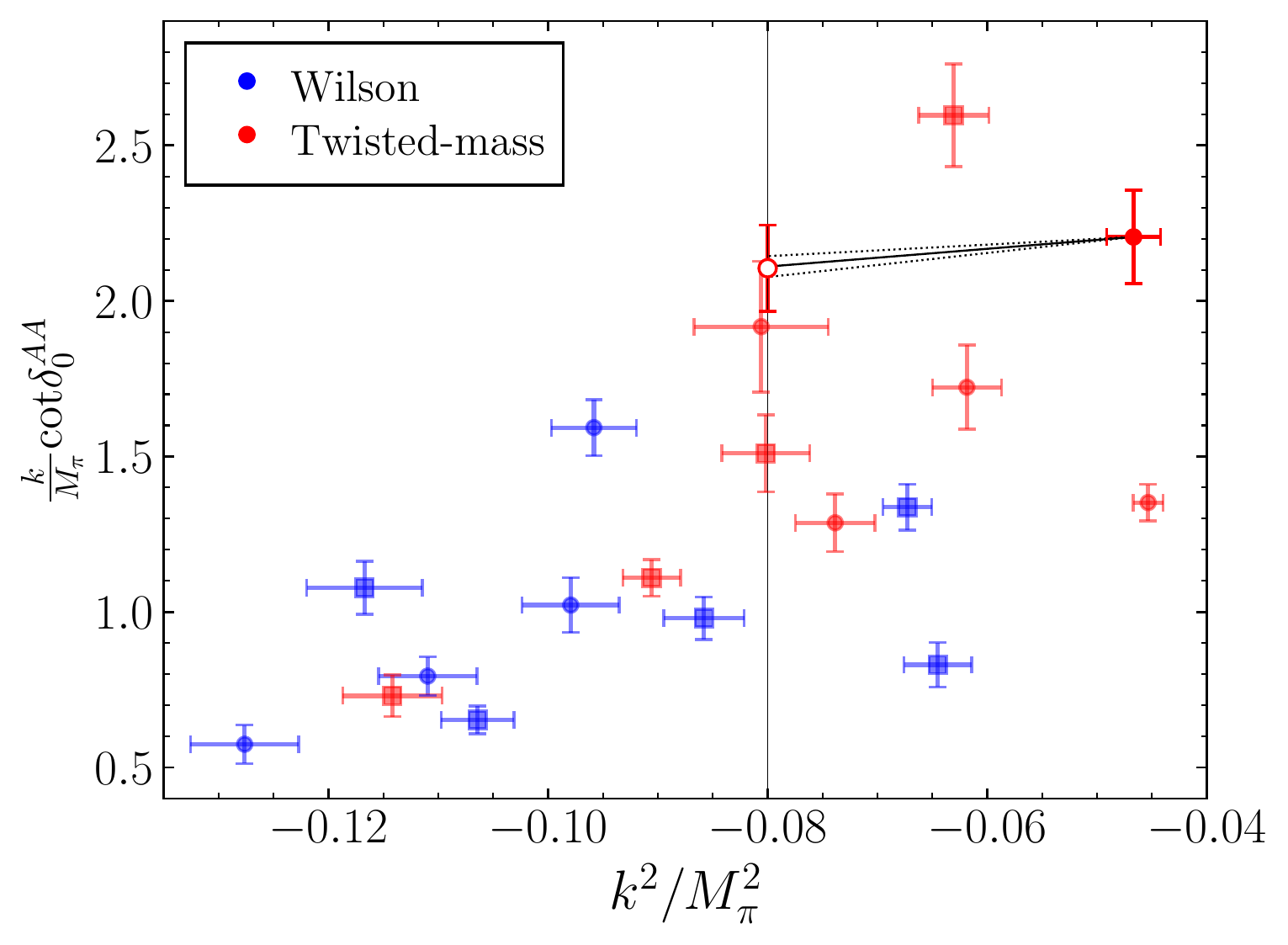}}
\caption{\label{fig:qExtrapolation} Results for the $s$-wave phase shift of the AA channel for two regularizations, as indicated in the figure legend. We graphically show how one of the points---corresponding to the 3A40 ensemble---is extrapolated to $k_\text{ref}^2=-0.08M_\pi^2$, using the ERE and a prior for the effective range based on ChPT, $M_\pi^2a_0^{AA}r_0^{AA}\in[-5,-1]$, as explained in the main text. Dotted lines illustrate the error introduced by the width of this interval.}
\end{figure}

The procedure is as follows. We first extrapolate each ensemble to $k_\text{ref}^2=-0.08M_\pi^2$, using Eq.~(\ref{eq:ERE}) with a prior for the effective range, $M_\pi^2r_0^{AA}a_0^{AA}\in [-5,-1]$. This choice is inspired by the LO prediction from ChPT, Eq.~(\ref{eq:EffectiveRangeLO}). The width of the interval is chosen based on some recent results that show that this quantity may have sizeable higher order corrections~\cite{Blanton:2021llb}. An example of this extrapolation for one ensemble is shown in Fig.~\ref{fig:qExtrapolation}. Note that the errors introduced by the width of the prior, depicted by dotted lines, is negligible compared to the statistical one.

\begin{figure}[h!]
\centering
\subfigure%
[``A'' ensembles ($a=0.075$ fm)]{\includegraphics[width=0.505\textwidth,clip]{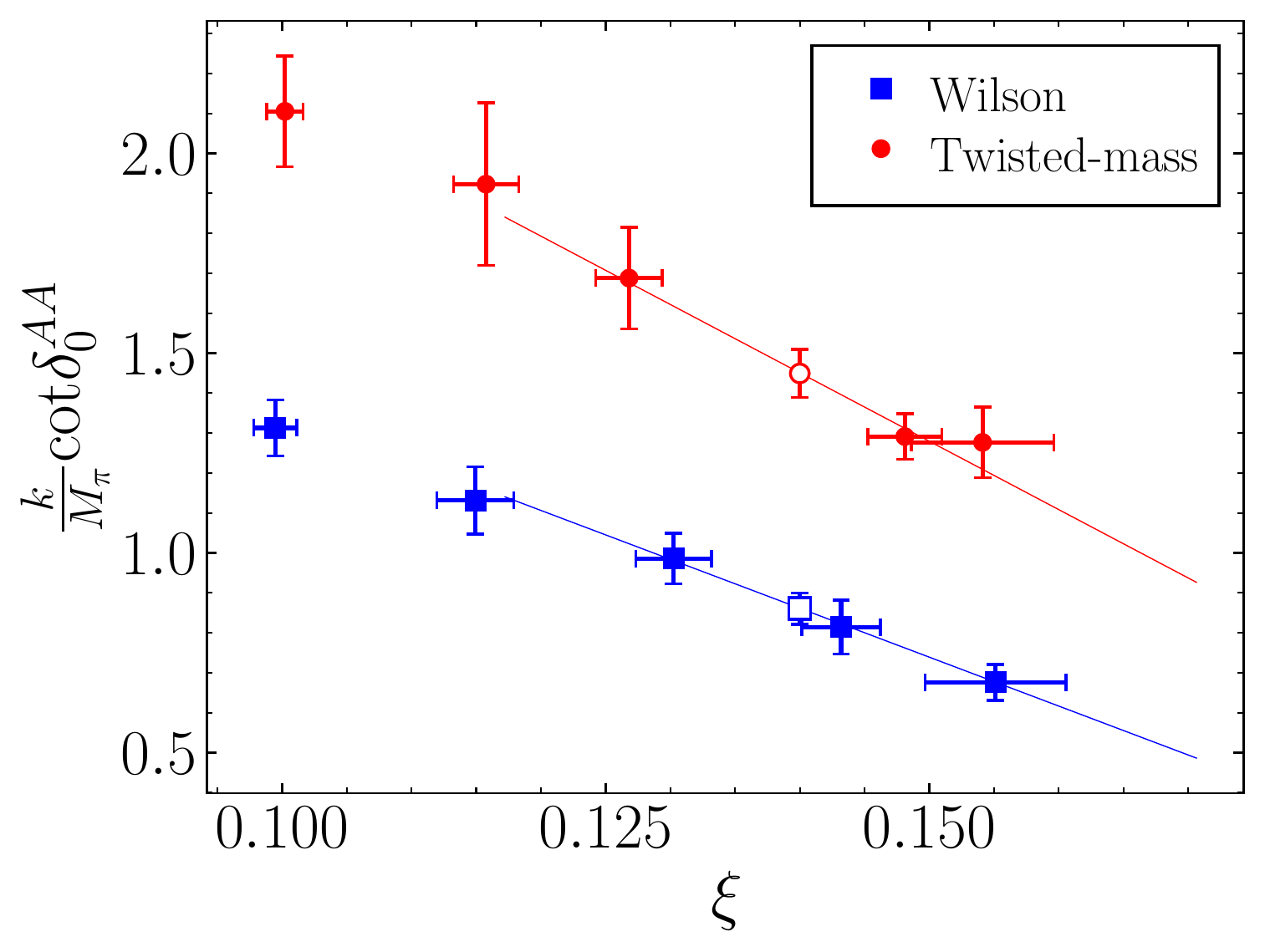}}\hfill
\subfigure%
[``B'' ensembles ($a=0.065$ fm)]{\includegraphics[width=0.485\textwidth,clip]{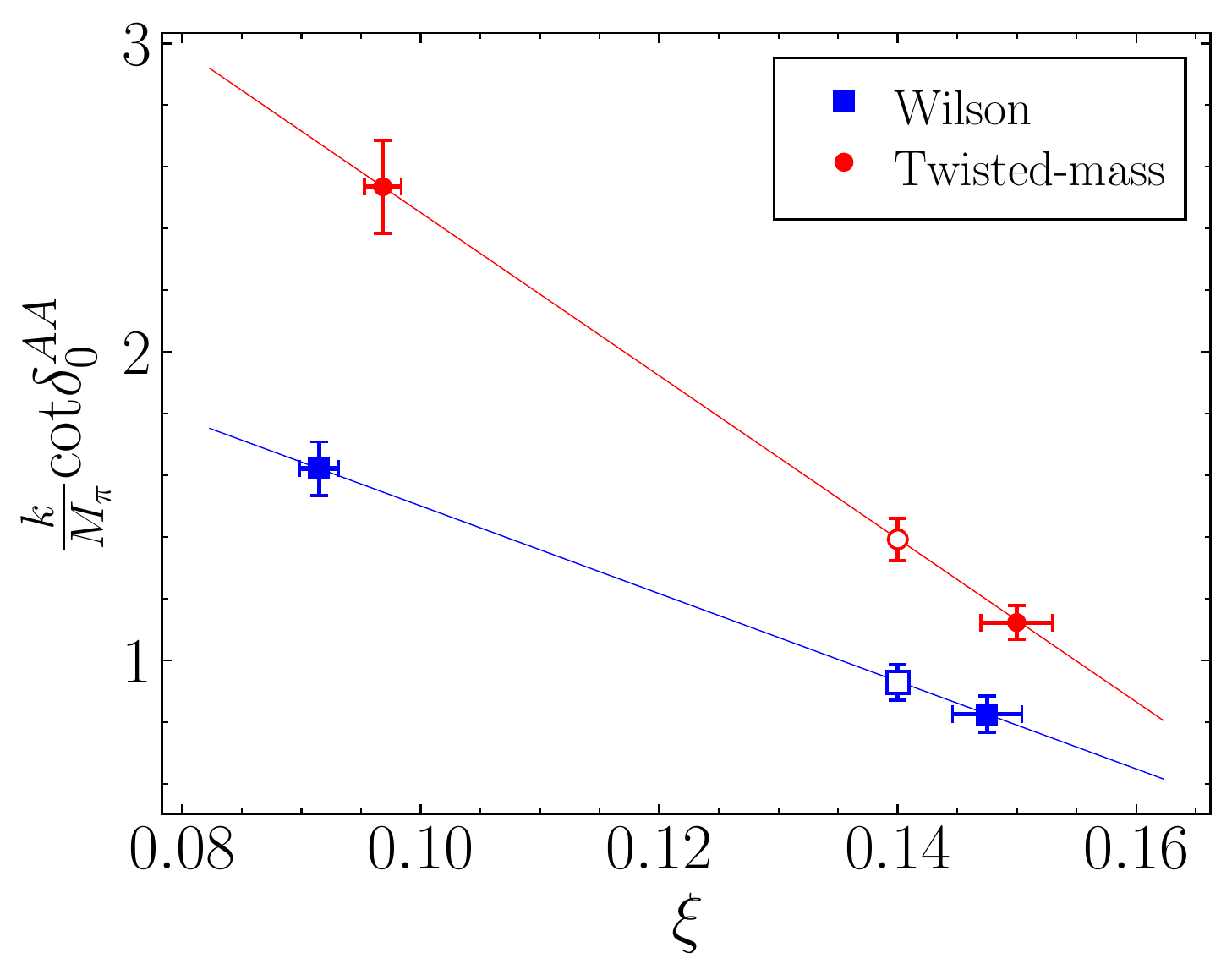}}
\subfigure%
[``C'' ensembles ($a=0.059$ fm)]{\includegraphics[width=0.505\textwidth,clip]{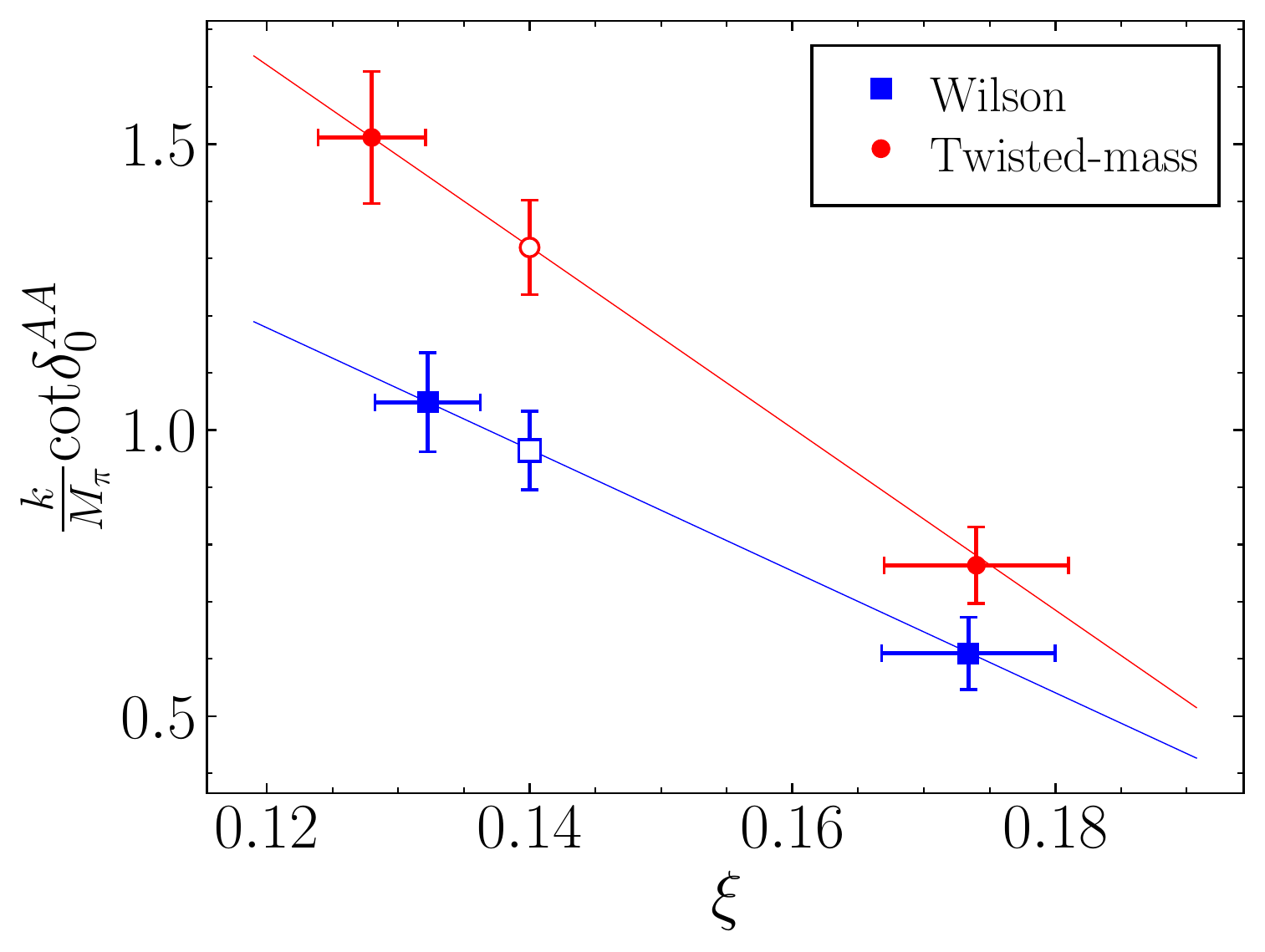}}
\caption{\label{fig:interpolation} Results for the linear interpolation of $({k}{M_\pi})\cot\delta_0$ to a fixed value $\xi_\text{ref}=0.14$, depicted as empty pointd, for the pure Wilson (blue) and mixed-action (red) setups, and for the three lattice spacings. In the coarser case, only three points are used in the interpolation.}
\end{figure}

\begin{figure}[h!]
\centering
\subfigure%
{\includegraphics[width=0.7\textwidth,clip]{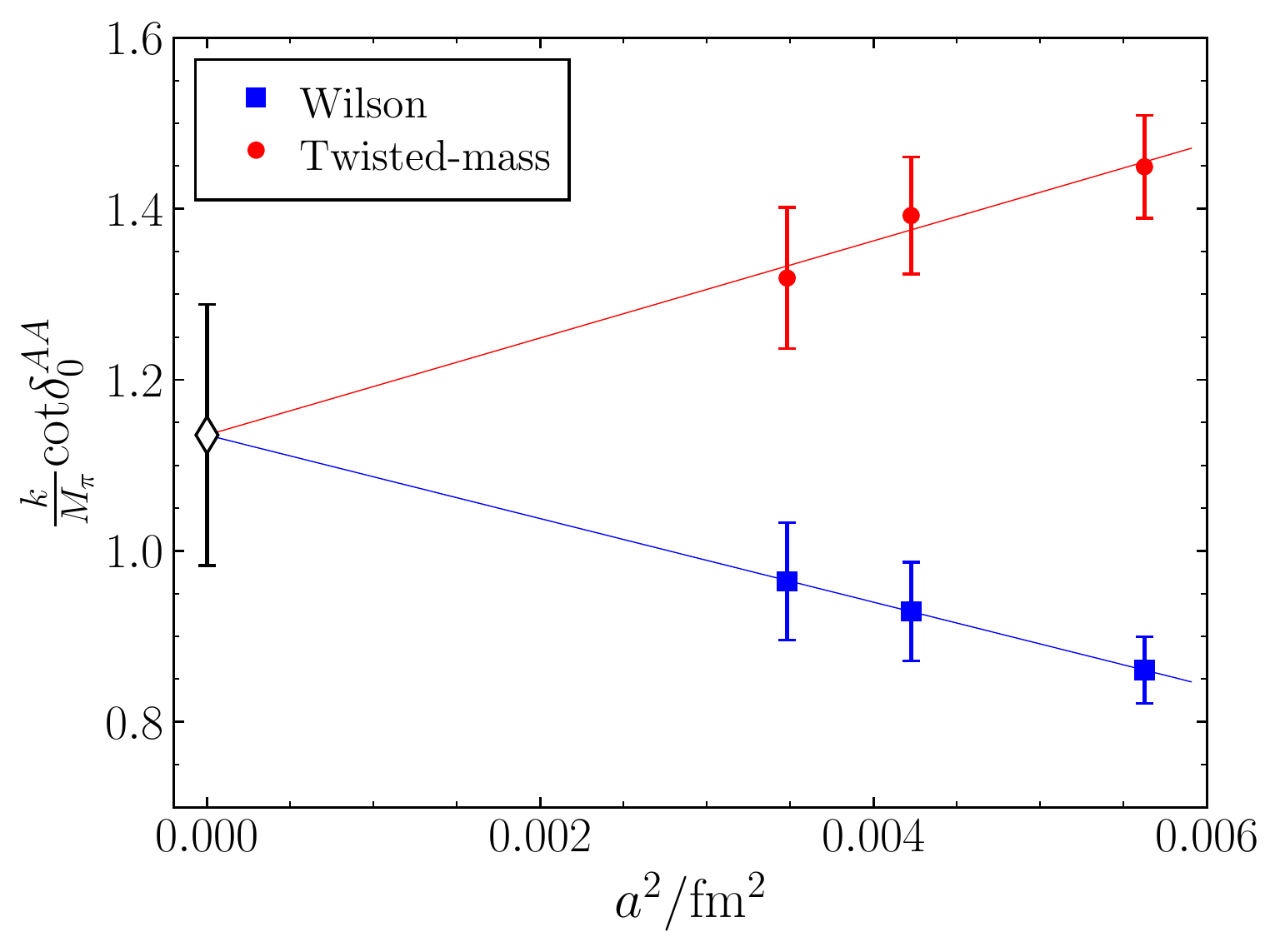}}
\caption{\label{fig:continuumextrapolation} Continuum extrapolation of $({k}/{M_\pi})\cot\delta_0$. Data from the pure Wilson (blue) and mixed-action (red) setups is fitted simultaneously with a constrained common continuum limit, depicted by an empty diamond.}
\end{figure}

Next, at fixed lattice spacing, we interpolate linearly to $\xi_\text{ref}=0.14$, as shown in  Fig.~\ref{fig:interpolation}. Finally, we use the results at the reference point to do a constrained linear continuum extrapolation using both regularizations, and taking into account correlations. The result is shown in Fig.~\ref{fig:continuumextrapolation}.  Note that we expect $\mathcal{O}(a)$ improvement in both regularizations, since we are using a non-vanishing value of $c_\text{sw}$. Our results are consistent with a regularization-independent continuum limit and $\mathcal{O}(a^2)$ scaling, as expected.

%Notice that the discretization errors have different sign for the two difference regularisations. 
Henceforth, we stick to the mixed-action setup and explicitly include discretization effects to analytical expressions. To do so we have studied a generalization of ChPT that considers the effects of a Wilson term as well as a twisted mass~\cite{Munster:2004wt, Sharpe:2004ny, Buchoff:2008hh}. We have not found in ChPT any hint pointing to larger discretization effects in the $AA$ channel compared to the $SS$ one. We have chosen to parameterize the dominant corrections as

\begin{equation}\label{eq:kcotDiscretization}
\mathcal{M}_{AA}^\text{latt}=\mathcal{M}_{AA}^\text{cont}+32\pi^2 a^2\xi W \longrightarrow \frac{k}{M_\pi}\cot\delta_0^\text{latt}=\frac{k}{M_\pi}\cot\delta_0^\text{cont}\left(1-\frac{32\pi^2 a^2 \xi W}{\mathcal{T}_{0,AA}^\text{LO}}\right),
\end{equation} %Important remark: The values we had fitted have an extra pi^-1
where $W$ a combination of new LECs that appear in Wilson ChPT, and is expected to scale as $\mathcal{O}(N_\text{c}^0)$,  $\mathcal{T}_{0,\text{LO}}^{AA}$ is the LO $s$-wave scattering amplitude for the $AA$ channel, as explained in App.~\ref{app:scatteringamplitude}, and we use ``cont'' (``latt'') to refere to continuum (lattice) quantities. More details about this result are explained in App.~\ref{app:WilsonChPT}.

From the continuum extrapolation, it is possible to determine a value 
\begin{equation}\label{eq:Wfromcontextrap}
W=-42(29) \text{ fm}^{-2}.
\end{equation}
 This result will be used as an additional input when matching the $AA$ channel results to ChPT.
%Note we have not studied effects related to the mixed-action. Due to their degree of complexity, these lie out of the scope of this work. \red{No entiendo muy bien que quiere decir estas dos ultimas frases}

\section{Fits to Chiral Perturbation Theory} \label{sec:fits}

We now match the lattice results for $\Delta E_R$ to ChPT predictions in Sec.~\ref{sec:ChPT} and App.~\ref{app:scatteringamplitude}. This way we extract the relevant LECs and study their large $N_\text{c}$ scaling.

\subsection{Fitting procedure}\label{sec:procedure}

In order to match our lattice results to ChPT, we need the prediction of $\Delta E_R$ in this effective theory. The starting point is the result for the scattering amplitudes that can be found in Ref.~\cite{Bijnens_2011} for the SU($N_\text{f}$) theory, and in Eqs.~(\ref{eq:ScattAmplUSS}) and~(\ref{eq:ScattAmplUAA}) for the U($N_\text{f}$) case. Following Ref.~\cite{Hernandez:2019qed}, we choose a convenient renormalization scale, which is related to $4\pi F_\pi$ but has no leading $N_\text{c}$ dependence,
\begin{equation}\label{eq:RenormScale}
\mu^2=\frac{3}{N_\text{c}}(4\pi F_\pi)^2. 
\end{equation}
Also, in the U($N_\text{f}$) theory, we take the splitting between $M_\pi^2$ and $M_{\eta'}^2$ to be
\begin{equation}\label{eq:etamass}
\frac{M_{\eta'}^2}{(4\pi F_\pi)^2}=\xi+\frac{a_0}{N_\text{c}^2},
\end{equation}
with $a_0\sim6.5$, as determined using Eq.~(\ref{eq:WittenVeneziano}) and the topological susceptibility from Ref.~\cite{Ce:2016awn}.

The results for the scattering amplitudes are then projected to an $s$-wave following Eq.~(\ref{eq:partialwave}), and the corresponding phase shift is computed using Eq.~(\ref{eq:phaseshift}). Discretization effects are also included for the $AA$ channel as explained in  Eq.~(\ref{eq:kcotDiscretization}).  Then, the L\"uscher equation [Eq.~(\ref{eq:FullLuscher})] is  numerically solved to obtain the finite-volume energies.\footnote{We use an efficient implementation for the generalized zeta function, introduced in Refs.~\cite{LuscherZetaGit,NPLQCD:2011htk}.} The energy shifts are finally determined as a function of $\xi$, $M_\pi L$ and the LECs.

Since we are working only for small values of the momentum,  the LECs that appear in the partial amplitudes multiplied by a power of the momentum---$L_R'$ and $L_R''$ in Eqs.~(\ref{eq:TNLOSS}) and~(\ref{eq:TNLOAA}) and $K_R'$ in Eqs.~(\ref{eq:ScattAmplUSS}) and~(\ref{eq:ScattAmplUAA})---are set to zero. This is justified, as we do not have enough information to constrain them.

Due to non-negligible correlations between $\Delta E_R$ and $\xi$, Ordinary Least Squares is not a good option for fitting. We instead use York Regression~\cite{doi:10.1119/1.1632486}, defining the $\chi^2$ function as
\begin{equation}\label{eq:chi2York}
\chi^2 = \min_{\delta_i}[\bm{R}^\intercal V^{-1} \bm{R}].
\end{equation}
where $\bm{R}$ is the data vector and $V$ is the corresponding covariance matrix. For a simple fit to a single channel, $\bm{R}$ is composed by a succession of tuples $(f(x_i+\delta_i)-y_i,\delta_i)$ corresponding to all the ensembles $i$ that are fitted. Here, $x_i=\xi_i$ and $y_i=\Delta E_{R,i}$ are our lattice results, and $f$ the ChPT prediction of the energy shift, computed as explained above. In this case $V$ is a block-diagonal matrix, due to the statistical independence of the different ensembles.

By contrast, when we fit discretization effects in the $AA$ channel, additional correlations must be included. This is due inclusion in $\bm{R}$ of the value of $W$ from Eq.~(\ref{eq:Wfromcontextrap}), and so, $V$ is no longer block-diagonal. Similarly when we consider a simultaneous fit to both channels, the tuples need to be enlarged, 
\begin{equation}
(f^{SS}(x_i+\delta_i)-y_i^{I=2},f^{AA}(x_i+\delta_i)-y_i^{AA}, \delta_i).
\end{equation}

Note that, even though the energy shifts depend on both $\xi$ and $M_\pi L$, only uncertainties in the former are considered. The reason is that relative errors in the latter quantity are much smaller, so we expect them to have a insignificant effect. We have checked this explicitly for some cases and found this to be true.

\subsection{Fits at fixed $N_\text{c}$} \label{sec:fixedNc}

 We first work at fixed $N_\text{c}$ and fit the energy shifts of both channels to SU(4) and U(4) ChPT, following the procedure described above. In all cases, we  determine the values of $L_R$ in Eqs.~(\ref{eq:TNLOSS}) and~(\ref{eq:TNLOAA}) from the fits. Additionally, we determine $K_R$, as defined in Eqs.~(\ref{eq:ScattAmplUSS}) and~(\ref{eq:ScattAmplUAA}), when we fit to the U(4) theory. Finally, for the $AA$-channel, we also fit the  parameter $W$ that accounts for discretization effects, as introduced in Eq.~(\ref{eq:kcotDiscretization}).
 
The results for the fits are shown in Tables~\ref{tab:I2fixedNc} and~\ref{tab:AAfixedNc} for the $SS$ and $AA$ channels, respectively. In the former case,  ChPT at the  order considered cannot describe the behaviour of the lattice results for the heaviest masses. We have thus not included ensembles with $\xi \gtrsim 0.14$ in the fit. In addition, as we are using the $W$ determination from the continuum extrapolation as an additional input when fitting the $AA$ channel results, the number of degrees of freedom (dof) is increased by one unit in this case.

\begin{figure}[h!]
   \centering
   \subfigure[SU(4) ChPT]%
             {\includegraphics[width=0.48\textwidth,clip]{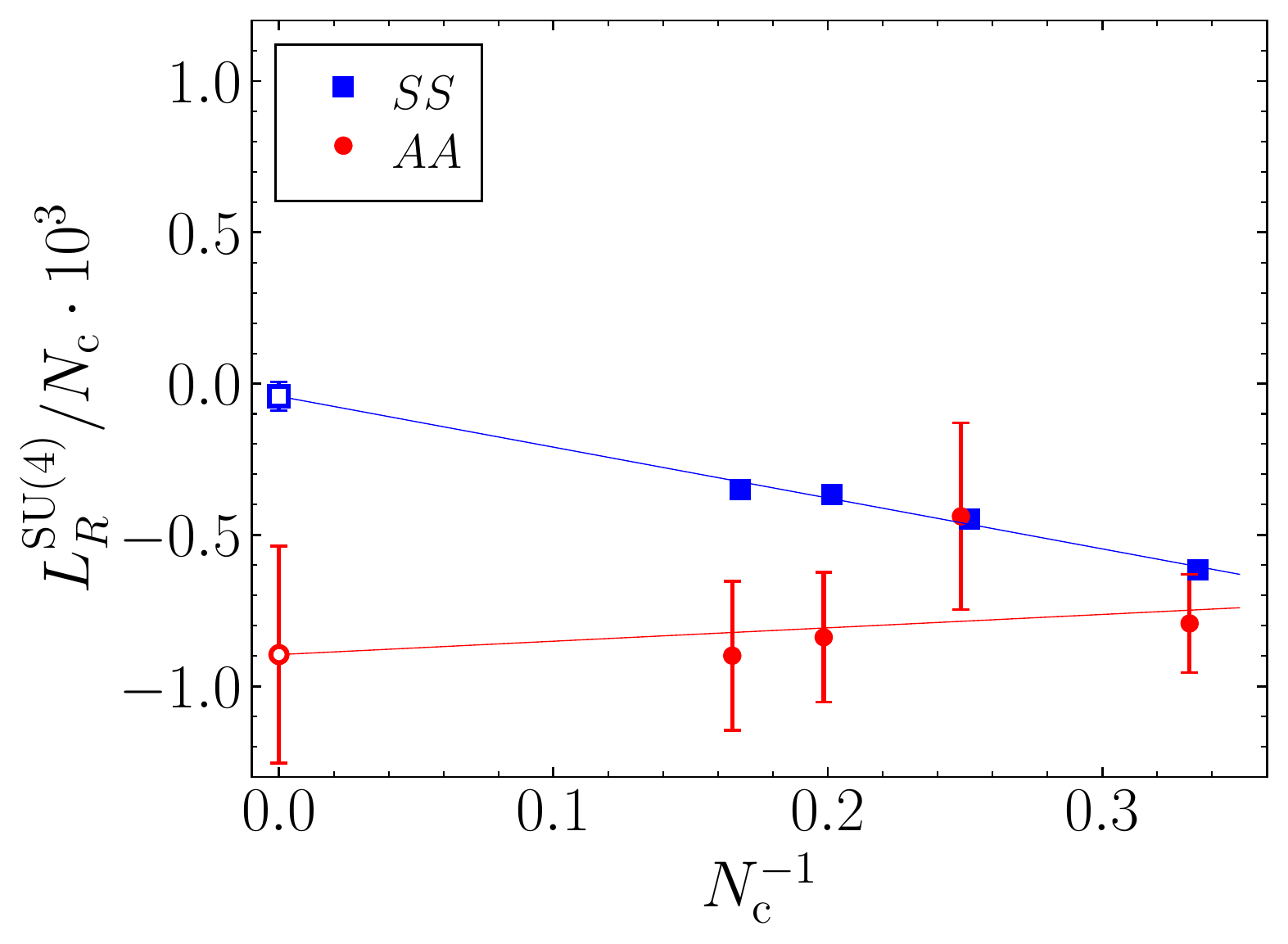}}
   \subfigure[U(4) ChPT]%
             {\includegraphics[width=0.48\textwidth,clip]{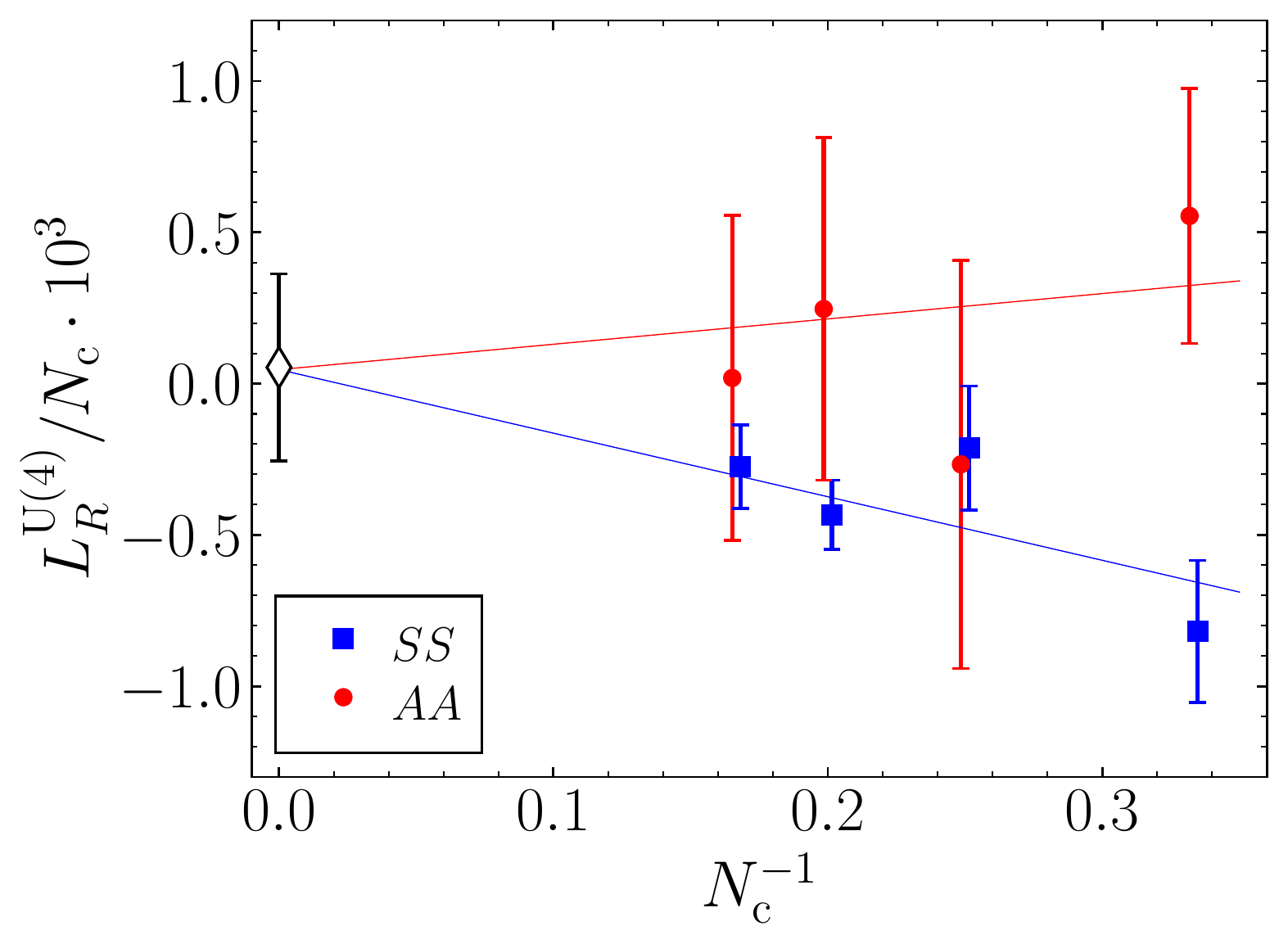}}
   \caption{Results for the fits to ChPT at fixed $N_\text{c}$ for the $SS$ (blue squares, Table~\ref{tab:I2fixedNc}) and $AA$ (red circles, Table~\ref{tab:AAfixedNc}) channels. The lines are the best fit to Eq.~(\ref{eq:LECsExpansion}), and empty points represent the large $N_\text{c}$ limit. In the case of the U(4) theory, we have imposed a common limit for the fit.
   }
   \label{fig:LSUandU}
\end{figure}

The  $N_\text{c}$ scaling of the fitted values of $L_{SS}$ and $L_{AA}$ is shown in Fig.~\ref{fig:LSUandU}.  It is well described by the expected  leading and subleading $N_\text{c}$ dependence. However, only in the U(4) theory the leading term is found to be consistent in the two channels---see  Eq.~(\ref{eq:LECsExpansion}).

\begin{figure}[h!]
   \centering
   \subfigure%
             {\includegraphics[width=0.7\textwidth,clip]{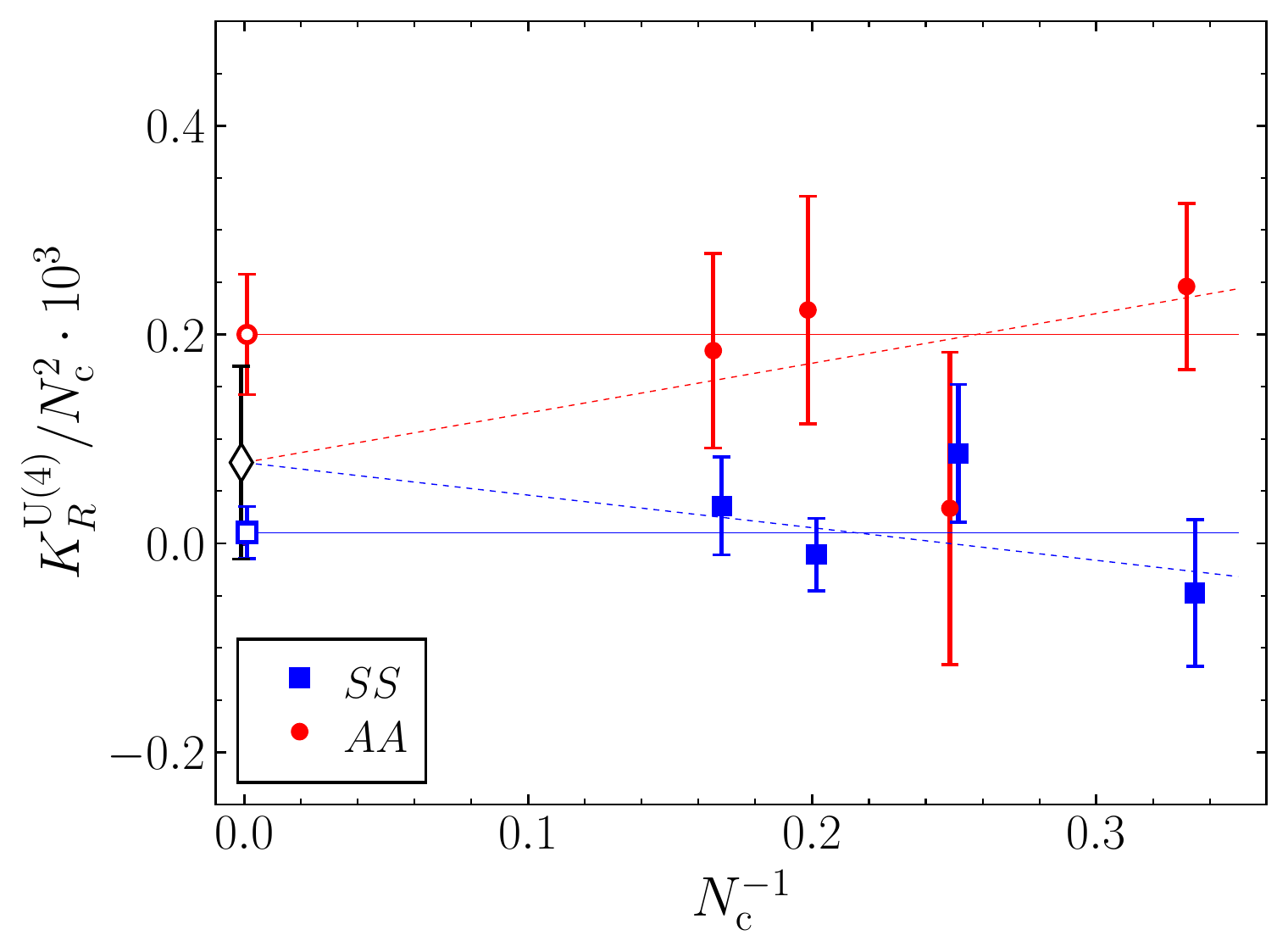}}
   \caption{Results for $K_R$ from the U(4) ChPT fits for the $SS$ (blue squares, Table~\ref{tab:I2fixedNc}) and $AA$ (red circles, Table~\ref{tab:AAfixedNc}) channels, together with two options for large $N_\text{c}$ extrapolations, represented by empty points. Solid lines represent fits to contant values, which are not consistent as expected from Eq~(\ref{eq:leadingK}). Dashed lines correspond to a linear fit with constrained dominant term and different subleading dependence.  }
   
   \label{fig:K}
\end{figure}

In Fig.~\ref{fig:K} we also study the $N_\text{c}$ scaling of $K_R$. At the order we are working, only the leading $N_\text{c}$ dependence should be included, which implies a constant value of the ratio shown in Fig.~\ref{fig:K}. However, the results found for both channels (solid lines) are not consistent, as would be expected from Eq.~(\ref{eq:leadingK}).  A common limit can be found from a constrained linear fit that includes both channels (dashed line), with different subleading $N_\text{c}$ corrections in $K_R$. %When fiting together all ensembles, nevertheless, we don't have enough information to constrain the subleading dependencies, so as a compromise, we have opted to keep two different leading-$N_\text{c}$ parameters for each channel, $K_R=K_R^{(0)}N_\text{c}^2$.

\begin{table}[h!]
\centering
\begin{tabular}{|c|c|c|c|c|c|}
\hline
 \multirow{2}{*}{$N_\text{c}$} & \multicolumn{2}{c|}{SU(4) ChPT} & \multicolumn{3}{c|}{U(4) ChPT} \\\cline{2-6}
 & $L_{SS}\cdot10^{3}$ & $\chi^2/\text{dof}$ & $L_{SS}\cdot10^{3}$ & $K_{SS}\cdot10^{3}$ & $\chi^2/\text{dof}$ \\   \hline
3 & $-1.85$(7) & 1.8/4 & $-2.5$(7) & $-0.4$(0.6) & 1.6/3\\
4 & $-1.79$(8) & 3.0/3 & $-0.9$(8) & 1.3(1.0) & 0.1/2\\
5 & $-1.83$(11) & 2.3/3 & $-2.2$(6) & $-0.3$(0.9) & 2.3/2\\
6 & $-2.10$(16) & 3.8/3 & $-1.7$(8) & 1.2(1.7) & 2.8/2\\ \hline 
\end{tabular}\hspace{1cm}
\caption{Results of fits of $\Delta E_{SS}$ to SU(4) and U(4) ChPT at fixed $N_\text{c}$. $N_\text{c}=3$ ensembles with $\xi \gtrsim 0.14$ are not fitted.}
\label{tab:I2fixedNc}\vspace{0.3cm}
\end{table}

\begin{table}[h!]
\centering
\begin{tabular}{|c|c|c|c|c|c|c|c|}
\hline
 \multirow{2}{*}{$N_\text{c}$}& \multicolumn{3}{c|}{SU(4) ChPT} & \multicolumn{4}{c|}{U(4) ChPT} \\ \cline{2-8} 
 & $L_{AA}\cdot10^{3}$ & $W$/fm$^{-2}$ & $\chi^2/\text{dof}$  & $L_{AA}\cdot10^{3}$ & $K_{AA}\cdot10^{3}$ & $W$/fm$^{-2}$ & $\chi^2/\text{dof}$ \\  \hline
3 & $-2.4$(5) & $-72$(17) & 23.4/8 & 1.7(1.3) & 2.2(0.7) & $-39$(23) & 12.8/7 \\ 
4 & $-1.8$(1.2) & $-45$(30) & 1.2/3 & $-1.1$(2.7) & 0.5(2.3) & $-41$(32) & 1.2/2 \\
5 & $-4.2$(1.1) & $-75$(22) & 8.5/3 & 1.2(2.8) & 5.6(2.7) & $-41$(32) & 3.9/2 \\
6 & $-5.4$(1.5) & $-72$(24) & 5.4/3 & 0.1(3.2) & 6.6(3.3) & $-39$(33) & 1.6/2 \\ \hline
\end{tabular}\hspace{1cm}
\caption{Results of fits $\Delta E_{AA}$ to SU(4) and  U(4) ChPT at fixed $N_\text{c}$.}
\label{tab:AAfixedNc}\vspace{0.2cm}
\end{table}

\subsection{Simultaneous chiral and $N_\text{c}$ fits} \label{sec:energyfit}

Given the results above, we are confident to perform global fits including all values of $N_\text{c}$. We have done this both for the SU(4) and the U(4) theories,  parametrizing $L_R$ with a leading and subleading $N_\text{c}$ terms. Regarding $K_R$ in the U(4) theory, we have opted to keep two different leading-$N_\text{c}$ parameters, one for each channel, since we do not have enough information to constrain  also subleading corrections,
\begin{equation}
K_R=K_R^{(0)}N_\text{c}^2+\mathcal{O}(N_\text{c}).
\end{equation}

%When fiting together all ensembles, nevertheless, we don't have enough information to constrain the subleading dependencies, so as a compromise, we have opted to keep two different leading-$N_\text{c}$ parameters for each channel, $K_R=K_R^{(0)}N_\text{c}^2$.

\begin{figure}[h!]
   \centering
      \subfigure%
             {\includegraphics[width=1\textwidth,clip]{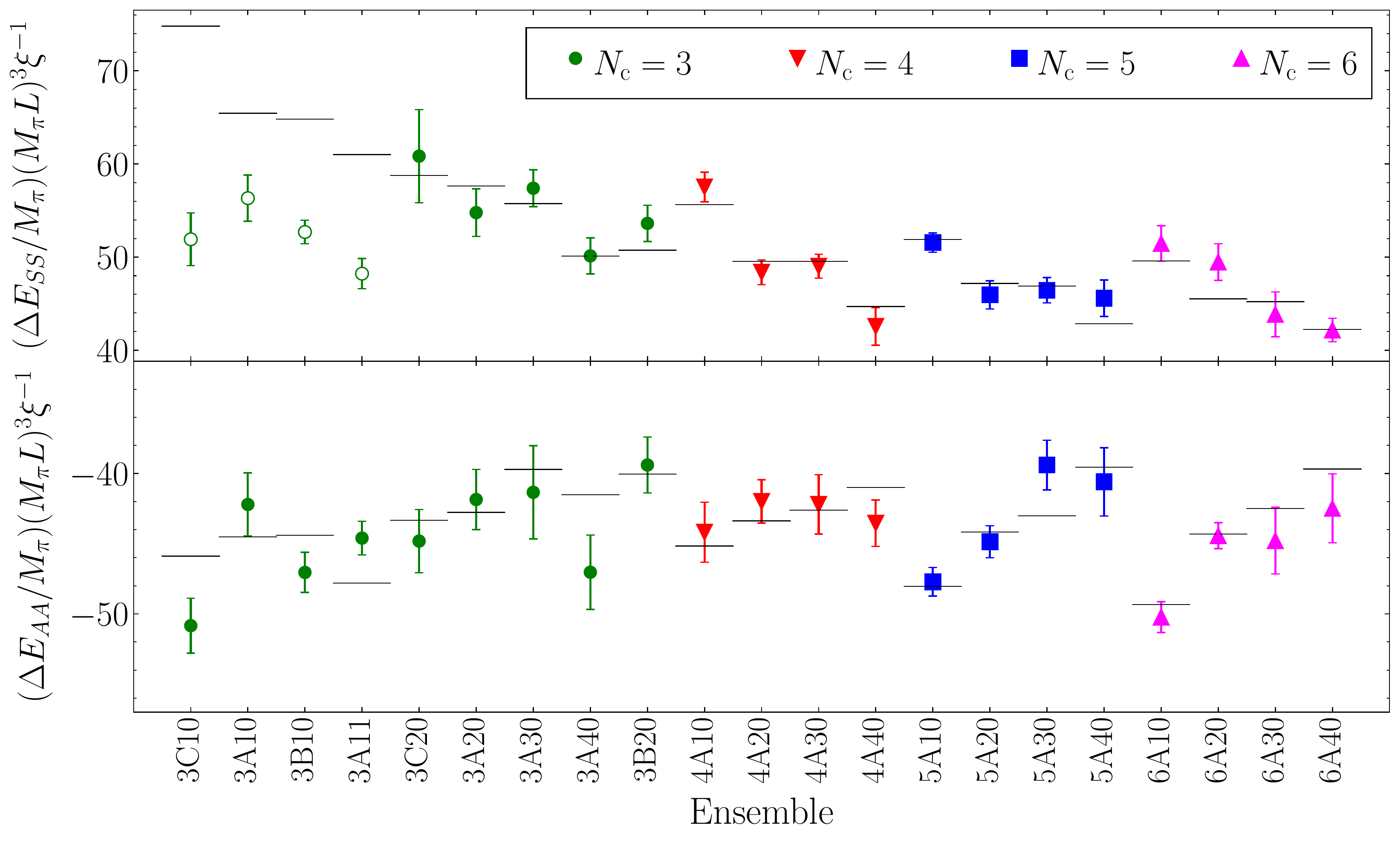}}
   \caption{Results for the simultaneous chiral and $N_\text{c}$ fit of the energy shifts for both $SS$ (top) and $AA$ (bottom) channels. Lattice results are depicted as points, with a different marker for different $N_\text{c}$ as explained in the plot legend. Empty points in the $SS$ channel are not fitted. Horizontal lines correspond to the best fit to U(4) ChPT, summarized in Table~\ref{tab:BothChannels}. The values are multiplied by $(M_\pi L)^3\xi^{-1}$ to eliminate leading chiral and $N_\text{c}$ dependence. }
   \label{fig:energyfit}
\end{figure}

First, we have performed fits to single channels for the two theories. The results are shown in Table~\ref{tab:SingleChannel}. The best fit results of $L^{(0)}$ are different for both channels when using the SU(4) theory, but they are compatible in the U(4) case, in agreement with the results from the previous section and theoretical expectations. This led us to perform a simultaneous fit of both channels to the U(4) theory with a constrained common value of $L^{(0)}$. The results are presented in Table~\ref{tab:BothChannels} and the quality of the fits is shown in Fig.~\ref{fig:energyfit}.

\begin{table}[h!]
\centering
\begin{tabular}{|c|c|c|c|c|c|c|}
\hline
Channel & Fit & $L^{(0)}\cdot10^{3}$ & $L_R^{(1)}\cdot10^{3}$ & $K_R^{(0)}\cdot10^{5}$ & $W$/fm$^{-2}$ & $\chi^2/\text{dof}$  \\  \hline
\multirow{2}{*}{$SS$} & SU(4) & $-0.04$(1.3) & $-1.70$(18) & $-$ & $-$ & 12.8/15  \\
 & U(4) & $-0.01$(7) & $-1.78$(20) & 1.2(2.5) & $-$ & 12.2/14 \\
 \hline
\multirow{2}{*}{$AA$} & SU(4) & $-1.22$(19) & 0.8(4) & $-$ & $-94$(15) & 38.5/19 \\
 & U(4) & $-0.1$(4) & 1.8(4) & 21(5) & $-32$(23) & 22.5/18 \\
\hline 
\end{tabular}
\caption{Results of fits to the energy shift in both $SS$ and $AA$ channels to SU(4) and U(4) ChPT, as indicated in the first column.}
\label{tab:SingleChannel}
\end{table}

\begin{table}[h!]
\centering
\begin{tabular}{|c|c|c|c|c|c|c|}
\hline
 $L^{(0)}\cdot10^{3}$ & $L^{(1)}_{SS}\cdot10^{3}$ & $L^{(1)}_{AA}\cdot10^{3}$ & $K_{SS}^{(0)}\cdot10^{5}$ & $K_{AA}^{(0)}\cdot10^{5}$ & $W$/fm$^{-2}$ & $\chi^2/\text{dof}$  \\  \hline
$-0.02$(8) & $-1.79$(19) & 1.7(4) & 0.8(2.2) & 22(3) & $-31$(9) & 35.6/33  \\

\hline 
\end{tabular}
\caption{Results of a global fit to the energy shift in U(4) ChPT. The goodness of the fit is ilustrated in Fig.~\ref{fig:energyfit}.}
\label{tab:BothChannels}
\end{table}

%\subsection{Discussion of the results}

\subsection{Comparison to previous literature}

 From the global fit  to U(4) ChPT, we observe that the leading $N_\text{c}$ dependence of the LEC combination that enters $\Delta E_{R}$ in both channels,  $L^{(0)}$, is anomalously small. The dominant terms in the LECs for the two channels of interest are therefore the ones subleading in $N_\text{c}$. If we recall Eq.~(\ref{eq:LRscaling}) and combine the results for the two channels, we can further obtain the $N_\text{c}$ and $N_\text{f}$ scaling of the following linear combinations of LECs:
\begin{equation}\label{eq:Nfscaling}
\begin{array}{rl}
L_0+L_3-L_5+L_8=&-0.02(8)N_\text{c}-0.01(5)N_\text{f}+\mathcal{O}(N_\text{c}^{-1}),\\
L_1+L_2-L_4+L_6=&-0.88(10)+\mathcal{O}(N_\text{c}^{-1}).
\end{array}
\end{equation}
The combination that is in principle leading, seems to be suppressed (both in the leading and subleading terms) with respect to the subleading one.

Using this scaling, we can compare our results to previous literature at $N_\text{c}=3$, which  exists for the SU($N_\text{f}$) theory and $N_\text{f} <4$. Thus, only the $SS$ channel is available for comparison. We begin from the U(4) results of Eq.~(\ref{eq:Nfscaling}), for which the $N_\text{f}$ scaling is known, and determine the value of $L_SS$ for the desired number of flavors. Then, we use the matching of Eq.~(\ref{eq:I2L1Matching}) to translate our result to the SU($N_\text{f}$) case. Finally, we change the renormalization scale to that used in the literature.\footnote{Note the running of the LECs differs between SU($N_\text{f}$)~\cite{Bijnens:2009qm} and U($N_\text{f}$) ChPT~\cite{Kaiser_2000}.} Our initial scale is $\mu=1.40(12)$ GeV, which we have determined averaging Eq.~(\ref{eq:RenormScale}) over all ensembles. The error of this quantity is a systematic one coming from subleading $N_\text{c}$ effects, while the statistical error is negligible.

%Since the difference between the results for the SU(4) and U(4) theories for this channel is small, as seen in  Table~\ref{tab:SingleChannel}, we will assume the results and $N_\text{f}$ scaling from Eq.~(\ref{eq:Nfscaling}).  We also need to change the renormalization scale of our result from our scale $\mu$ to the scale at which the results are cited in the literature. 

We first compare to $N_\text{f}=3$ results from ChPT fits to experiment. Setting the renormalization scale at the mass of the $\rho$ resonance, $M_\rho=0.77$ MeV, we obtain
\begin{equation}
L_{SS}^{N_\text{c}=3,\,N_\text{f}=3}(M_\rho)=-1.14(22)_\text{stat}(11)_\mu\cdot 10^{-3},
\end{equation}
where the first error comes from the fit and the second corresponds to the renormalization scale. This result agrees with the values reconstructed from previous literature,\footnote{The error of these results has been estimated adding in quadrature the independent errors from each LEC. It is expected to be reduced when correlations are taken into account.}
\begin{equation}
L_{I=2}^{N_\text{c}=3,\,N_\text{f}=3}=-0.9(1.5)\cdot 10^{-3} \quad\quad \text{ from Ref.~\cite{Bijnens:2014lea} (Table 1, 2}^\text{nd}\text{ column)},
\end{equation}
\begin{equation}
L_{I=2}^{N_\text{c}=3,\,N_\text{f}=3}=-0.7(3.2)\cdot 10^{-3} \quad\quad \text{ from Ref.~\cite{Gasser:1984gg}}.
\end{equation}

We also compare to lattice results for $N_\text{f}=2$ \cite{Aoki:2021kgd}. In this case, they are at a scale $\mu=\sqrt{2}F_\pi\approx130.2$ MeV and also a different quantity is quoted
\begin{equation}
l_{I=2}=512\pi^2L_{SS}^{N_\text{c}=3,N_\text{f}=2}.
\end{equation}
From our results, we obtain
\begin{equation}
l_{I=2}=4.3(1.2)_\text{stat}(0.5)_\mu,
\end{equation} 
which is agreement with
\begin{equation}
l_{I=2}=4.65(0.85)_\text{stat}(1.07)_\text{sys} \quad\quad \text{ from Ref.~\cite{Feng:2009ij}},
\end{equation}
\begin{equation}
l_{I=2}=3.79(0.61)_\text{stat}\left({}^{+1.34}_{-0.11} \right)_\text{sys} \quad\quad \text{ from Ref.~\cite{ETM:2015bzg}}.
\end{equation}
where the first reported error is statistic and the second is systematic.

Finally, we can compare with the results obtained in the context of Resonant Chiral Theory (RChT)~\cite{Ecker1989, Pich:2002, Ledwig:2014}. RChT assumes that the LECs at large $N_\text{c}$ are saturated by the contribution obtained from the integration of the low-lying resonances. Interestingly, the single-resonance approximation predicts the same value for both channels,
\begin{equation}
L_{R}^\text{RChT}=-0.07\cdot 10^{-3}N_\text{c}\quad\quad \text{Ref.~\cite{Ledwig:2014} (Table 1, 2}^\text{nd}\text{ column)},
\end{equation}
which agrees with our U(4) result at large $N_\text{c}$, but fails to reproduce the results at  $N_\text{c}=3$, where subleading effects dominate. %It differs from our result in less than a sigma. 

\subsection{A resonance pole in the $AA$ channel?}

In the case of the $AA$ channel, there is no direct comparison with the literature, since the channel only exists for $N_\text{f}\geq 4$. However,  its attractive nature makes it very interesting. An obvious question is whether there could be a resonance in it that might be interpreted as a tetraquark state. This channel has the correct flavor content, and spin-parity quantum numbers to couple to the recently observed  $X_0(2900)$ exotic state, which has been detected in the $D^- K^+$ channel at LHCb~\cite{LHCb:2020pxc,LHCb:2020bls}.

It is well known that ChPT violates unitarity as the CM energy approaches the mass of the lightest resonance. Some approaches have been used to increase the range of validity of ChPT, namely including the resonances in the effective theory~\cite{Harada:2003em, Harada:1996uz}, or using the inverse amplitude method~\cite{Truong:1988, DOBADO1990134, Truong:1991, Dobado:1993, Dobado:1997, GomezNicola:2001as, Pelaez:2004xp}. The latter restores unitarity by rewritting the $s$-wave partial amplitude as
\begin{equation}
\mathcal{T}_0=\frac{\left(\mathcal{T}_0^\text{LO}\right)^2}{\mathcal{T}_0^\text{LO}-\mathcal{T}_0^\text{NLO}},
\end{equation}
where $\mathcal{T}_0^\text{LO}$ and $\mathcal{T}_0^\text{NLO}$ are the LO and NLO parts of the amplitude, respectively, as explained in App.~\ref{app:scatteringamplitude}. It has been argued that this procedure provides a tool to constrain LECs from the observed resonances, and inversely, to predict the presence of resonances, such as the $\sigma$, from the LECs as determined from phenomenological fits \cite{Dobado:1993,Dobado:1997,GomezNicola:2001as,Pelaez:2004xp}.

\begin{figure}[h!]
   \centering
   \subfigure%
             {\includegraphics[width=0.7\textwidth,clip]{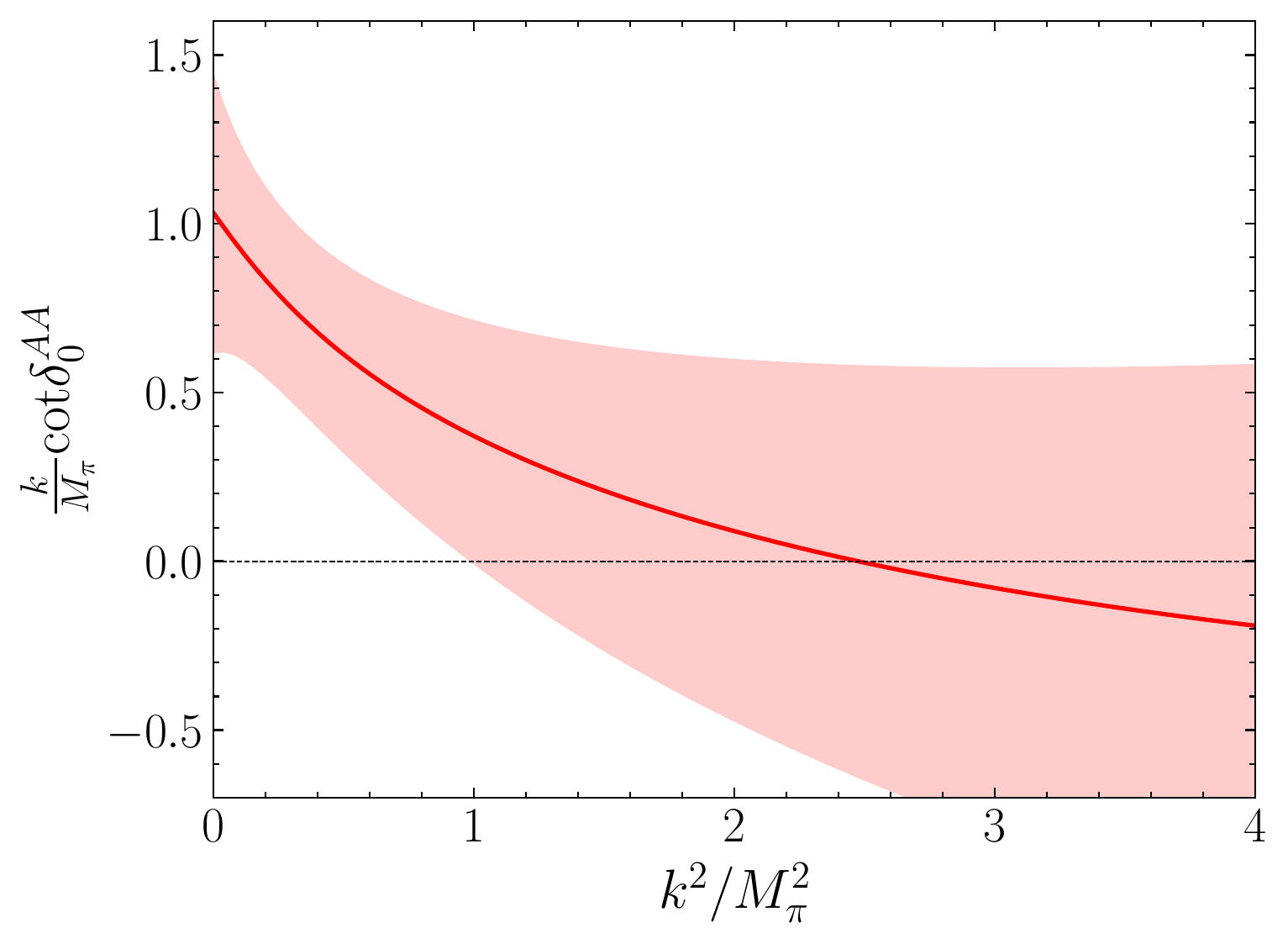}}
   \caption{$s$-wave phase shift for the $AA$ channel in U(4) ChPT for $\xi=0.1$ and $N_\text{c}=3$, computed using the inverse amplitude method. We have set the LECs which are not proportional to momenta in Eqs.~(\ref{eq:TNLOAA}) and~(\ref{eq:ScattAmplUAA}) to the fitted values, and restricted the rest to be of the same order, as explained in the main text. We observe that the curve may change sign, indicating the possible existence of a tetraquark resonance.}
   
   \label{fig:IAM}
\end{figure}

We can apply this approach to the $AA$ channel looking for the possible presence of a resonance.  In Fig.~\ref{fig:IAM}, we represent the $s$-wave phase shift for this channel in the U(4) theory for $\xi=0.1$ and $N_\text{c}=3$, as a function of the CM momentum. We use our best fit result for $L_{AA}$ and $K_{AA}$, while  we vary the LECs combinations $L_{AA}'$ and $L_{AA}''$---as defined in Eqs.~(\ref{eq:TNLOAA})  and (\ref{eq:ScattAmplUAA})---from zero (solid line) to $\pm L_{AA}$ (shaded band). $K_{AA}'$ is set to zero.  We conclude from this analysis that it is plausible that the curve crosses the positive $x$-axis from above, which would indicate the presence of a resonance.

%The  rest have been set to be of some arbitrary value of the correct order  of magnitude. 
%\red{Esto no se entiende: and their effect is taken into account in the uncertainty.}

If such a resonance exits, an interesting question is whether one could consider it a tetraquark state, as opposed to a meson-meson molecule. 
 Weinberg's criterium~\cite{Weinberg:1965, Matuschek:2020gqe} relates the probability of the state being a pure tetraquark, rather than a molecular state, to its field renormalization, that can be obtained in terms of the scattering length, $a_0$, and the effective range, $r_0$, of the channel as
\begin{equation}
Z=1-\sqrt{\frac{1}{1-\frac{2r_0}{a_0}}}.
\end{equation}
Setting $\xi=0.1$, $N_\text{c}=3$ and  using the LO results from Eq. (\ref{eq:EffectiveRangeLO}) for the effective range, and the NNLO U(4) prediction for the scattering length given in Eq.~(\ref{eq:AAUScatteringLength}),\footnote{We use the fit results for the LECs and the renormalization scale from Eq.~(\ref{eq:RenormScale}).} we get a value $M_\pi a_0^{AA}=0.51(4)$, and $Z\approx0.8$.  A value fo $Z$ so close to one indicates that this state, if it exists, would  more likely be a tetraquark rather than a meson molecule.  In future work, we plan to address the possible existence of this resonance, by studying the phase shift at different values of the CM momenta.

\section{Conclusions} \label{sec:conclusions}
In this work we have studied two-pion scattering in the context of the large $N_\text{c}$ limit of QCD with $N_\text{f}=4$. We have focused on two scattering channels that correspond to irreducible representations of SU(4). The first channel, the fully symmetric $SS$, is equivalent to the standard QCD isospin-2. The second channel is antisymmetric in both quarks and antiquarks, and is denoted as $AA$. It is an attractive channel that includes states with four open flavors and is only present in $N_\text{f} \geq 4$.

We have generated lattice simulations with HiRep~\cite{DelDebbio:2008zf,DelDebbio:2009fd} using $\mathcal{O}(a)$ improved Wilson fermions, and the Iwasaki gauge action. Most of the ensembles had already been used in previous work~\cite{Hernandez:2019qed,Donini:2020qfu}, and we have simulated a finer lattice spacing to gain further insight on discretization effects. In our setup, we use two different valence Dirac operators: Wilson (unitary setup) and twisted-mass (mixed action). We have then determined the two-pion energy levels in all ensembles and regularizations from a fit to the two-pion correlation functions. The technical details of the lattice setup can be found in Sec.~\ref{sec:setup}.

A possibility to analyze the energy levels is the use of the $1/L$ expansion of the ground state---the so-called threshold expansion. While this works nicely for the $SS$ channel, we have found that the converge of this expansion is poorer for the attractive $AA$ channel. Therefore, we decided to use the full L\"uscher formalism that is non-perturbative in $1/L$. Since we are working with mesons near threshold, we only need to include $s$-wave interactions.

Whereas both regularizations in the valence sector should agree in the continuum limit, that may not be the case at finite lattice spacing. Indeed, we found that even at the level of the energy shifts, the two regularizations show significant cutoff effects in the $AA$ channel. Our continuum extrapolation at $N_\text{c}=3$ indicated that cutoff effects account for a $\sim 20\%$ impact in the phase shift at $a=0.075$ fm. This result, along with Ref.~\cite{Green:2021qol}, is one of the few evidences of significant discretization effects in hadron-hadron scattering.

We finally analyze our data matching the two-pion amplitude to Chiral Perturbation Theory. In this case, the correct effective description is U($N_\text{f}$) ChPT, since the flavor singlet meson becomes light at large $N_\text{c}$. Since no results for pion-pion scattering were available in the literature for U$(N_\text{f})$ ChPT, we have computed them in Sec.~\ref{sec:ChPT}---see also App.~\ref{app:scatteringamplitude}. We find that the leading coefficients in the $N_\text{c}$ scaling of the LECs that enter these observables are parametrically small, and that the value at $N_\text{c}=3$ is dominated by the subleading coefficient in the $1/N_\text{c}$ expansion. Interestingly, within the large $N_\text{c}$ limit, we also obtain the value of the LECs at $N_\text{c}=3$ and $N_\text{f}=2,3$, that agrees nicely with previous literature. In addition, we comment on the comparison of our large $N_\text{c}$ result to the LECs predicted in resonant ChPT: we find good agreement with the leading $N_\text{c}$ result.

The higher-energy behaviour of the $AA$ channel, and the interplay with the large $N_\text{c}$ limit is a particularly interesting avenue for future work. At threshold, it is an attractive channel, which could lead to the presence of a resonance at higher values of the centre-of-mass momentum. Given the presence of four open flavors in this channel, a resonance would be a candidate for a tetraquark state. Indeed, the recently found $X_0(2900)$ scalar resonance~\cite{LHCb:2020pxc,LHCb:2020bls} has the correct flavor content to overlap with this channel. In addition, a rough estimate based on the Inverse Amplitude Method  in Fig.~\ref{fig:IAM} indicates that it is plausible that such a resonance exists.

\acknowledgments

We thank Carlos Pena for useful discussions.

We acknowledge the support provided by the Generalitat Valenciana grant PROMETEO/2019/083, the European project H2020-MSCA-ITN-2019//860881-HIDDeN, and the AEI project  PID2020-113644GB-I00/AEI/10.13039/501100011033. JBB is also supported by the Spanish grant FPU19/04326 of MU.  FRL acknowledges funding from the EU Horizon 2020 research and innovation program under the Marie Sk{\l}odowska-Curie grant agreement No. 713673 and ``La Caixa'' Foundation (ID 100010434, LCF/BQ/IN17/11620044). FRL has also received financial support from Generalitat Valenciana through the plan GenT program (CIDEGENT/2019/040). The work of FRL has been supported in part by the U.S.~Department of Energy, Office of Science, Office of Nuclear Physics, under grant Contract Numbers DE-SC0011090 and DE-SC0021006.

We thank Mare Nostrum 4 (BSC), Finis Terrae II (CESGA), Cal\'endula (SCAYLE), Tirant 3 (UV) and Lluis Vives (Servei d’Inform\`atica UV) for the computing time provided.

\newpage
\appendix

\section{SU($N_\text{f}$) and U($N_\text{f}$) ChPT scattering amplitudes} \label{app:scatteringamplitude}

In order to match the lattice results to ChPT, we need to be able to relate the two-pion infinite-volume scattering amplitude to the corresponding finite-volume energy spectra. This is done using Lüscher's formalism, which reduces to Eq.~(\ref{eq:FullLuscher}) in the case of $s$-wave scattering. The $s$-wave phase shift required in this expression can be obtained from the $s$-wave projected amplitude. In general, the $\ell$ partial wave is defined as
\begin{equation}\label{eq:partialwave}
\mathcal{T}_{\ell,R}=\frac{2\ell+1}{2}\int_{-1}^{1}\mathcal{M}_R(s,t,u)\text{P}_\ell(\cos\theta)\text{d}\cos\theta,
\end{equation}
where $s$, $t$ and $u$ are Mandelstam variables normalized by the mass of the pion, $\text{P}_\ell$ are the Legendre polynomials and $\theta$ is the scattering angle in the CM frame. Here, $\ell=0$ denotes the $s$-wave result. The corresponding phase shift is then obtained as
\begin{equation}\label{eq:phaseshift}
k\cot\delta_0^{R}=16\pi\sqrt{s}\,\text{Re}\left(\frac{1}{\mathcal{T}_{0,R}}\right).
\end{equation} 

In the case of SU($N_\text{f}$) ChPT, the scattering amplitudes have been computed up to NNLO in Ref.~\cite{Bijnens_2011}. One can project them to $s$-wave and compute the phase shift. Here, we quote the $s$-wave projection up to NLO, using $x_2=M_\pi/F_\pi$ and $q^2=k^2/M_\pi^2$ as shorthand notation,
\begin{equation}
\mathcal{T}_{0,SS}^{\text{LO}}=-\mathcal{T}_{0,AA}^{\text{LO}}=-x_2(2+4q^2),
\end{equation}
\begin{equation}\label{eq:TNLOSS}
\def\arraystretch{1.7}
\begin{array}{rl}
\frac{\mathcal{T}_{0,SS}^{\text{NLO}}}{x_2^2}& = 32L_{SS}+32q^2L_{SS}'+\frac{128}{3}q^4L_{SS}'' \\
& +\frac{1}{4\pi^2}\left[-1-\frac{1}{N_\text{f}^2}+\frac{1}{N_\text{f}}+q^2\left(-3-\frac{N_\text{f}}{18}\right)+q^4\left(-\frac{10}{3}-\frac{11N_\text{f}}{27}\right)\right]\\
& +\frac{1}{4\pi^2}\left[-1-\frac{1}{N_\text{f}^2}+\frac{1}{N_\text{f}}+q^2\left(-3-\frac{N_\text{f}}{6}\right)+q^4\left(-\frac{10}{3}-\frac{5N_\text{f}}{9}\right)\right]\ln{\frac{M_\pi^2}{\mu^2}} \\
& + \left(2+8q^2+8q^4\right)J(s) + F_{SS}(q^2), 
\end{array}
\end{equation}
\begin{equation}\label{eq:TNLOAA}
\def\arraystretch{1.7}
\begin{array}{rl}
\frac{\mathcal{T}_{0,AA}^{\text{NLO}}}{x_2^2}& = -32L_{AA}+32q^2L_{AA}'+\frac{128}{3}q^4L_{AA}'' \\
& +\frac{1}{4\pi^2}\left[-1-\frac{1}{N_\text{f}^2}-\frac{1}{N_\text{f}}+q^2\left(-3+\frac{N_\text{f}}{18}\right)+q^4\left(-\frac{10}{3}+\frac{11N_\text{f}}{27}\right)\right]\\
& +\frac{1}{4\pi^2}\left[-1-\frac{1}{N_\text{f}^2}-\frac{1}{N_\text{f}}+q^2\left(-3+\frac{N_\text{f}}{6}\right)+q^4\left(-\frac{10}{3}+\frac{5N_\text{f}}{9}\right)\right]\ln{\frac{M_\pi^2}{\mu^2}} \\
& + \left(2+8q^2+8q^4\right)J(s) + F_{AA}(q^2). 
\end{array}
\end{equation}
Here $F_R(q^2)$ are integrals that need to be numerically evaluated, 
\begin{equation}
\def\arraystretch{1.7}
\begin{array}{rl}
F_{SS}(q^2)=&\int_{-1}^1\left[1+\frac{2}{N_\text{f}^2}-\frac{2}{N_\text{f}}+\frac{2N_\text{f}}{3}+q^2\left(2+\frac{4N_\text{f}}{3}-2x\right) \right.\\&\left. + q^4 \left(2+N_\text{f}-4x-\frac{4N_\text{f}x}{3}+2x^2+\frac{N_\text{f}x^2}{3}\right)\right] J(t(x))\text{d}x,
\end{array}
\end{equation}
\begin{equation}
\def\arraystretch{1.7}
\begin{array}{rl}
F_{AA}(q^2)=&\int_{-1}^1\left[1+\frac{2}{N_\text{f}^2}+\frac{2}{N_\text{f}}-\frac{2N_\text{f}}{3}+q^2\left(2-\frac{4N_\text{f}}{3}-2x\right) \right.\\&\left. + q^4 \left(2-N_\text{f}-4x+\frac{4N_\text{f}x}{3}+2x^2-\frac{N_\text{f}x^2}{3}\right)\right] J(t(x))\text{d}x,
\end{array}
\end{equation}
 $J(x)$ is a scalar loop integral that can be found in App.~C of Ref.~\cite{Scherer:2002tk}, and $s=4(1+q^2)$ and $t(x)=-2q^2(1-x)$. Also new linear combinations of LECs have been defined,
\begin{equation}
\def\arraystretch{1.5}
\begin{array}{ll}
L_{SS}'=4L_0+4L_1+6L_2+2L_3-2L_4-L_5, &\quad  L_{SS}''=3L_0+2L_1+4L_2+L_3,\\
L_{AA}'=-4L_0+4L_1+6L_2-2L_3-2L_4+L_5, &\quad  L_{AA}''=-3L_0+2L_1+4L_2-L_3.\\
\end{array}
\end{equation}  
The imaginary part of these NLO partial amplitudes, originated from $J(x)$ when $x>4$ and required to compute the phase shift, can be related to the LO amplitude by the optical theorem,
\begin{equation}\label{eq:OpticalTheorem}
\text{Im}\mathcal{T}_{0,R}^\text{NLO}=\frac{\left(\mathcal{T}_{0,R}^\text{LO}\right)^2}{32\pi}\frac{|\bm{q}|}{\sqrt{1+q^2}}.
\end{equation}

Regarding the U($N_\text{f}$) theory, there are no results available in the literature to the best of our knowledge. We have computed the scattering amplitudes in the two channels of interest to NNLO in the counting of Eq.~(\ref{eq:UcountingChPT}), and used them to predict the lattice energy spectra, as explained in Sec.~\ref{sec:procedure} of the main text. The additional contributions needed to compute the scattering amplitude are explained in Eq.~(\ref{eq:additionalUdiagrams}). We obtain
\begin{equation}\label{eq:ScattAmplUSS}
\begin{array}{rl}
 \mathcal{M}_{SS}^{\text{U}(N_\text{f})}=&\mathcal{M}_{SS}^{\text{SU}(N_\text{f})} - \frac{2M_\pi^4}{F_\pi^4 N_\text{f}}\left(1-\frac{2}{N_\text{f}}\right)B_1(t)-\frac{2M_\pi^4}{F_\pi^4 N_\text{f}^2}B_2(t)  + (t\leftrightarrow u) \\
& + \left(\frac{M_\pi^2}{F_\pi^2}\right)^3\left[K_{SS}+q^2K_{SS}'\right],
\end{array}
\end{equation}
\begin{equation}\label{eq:ScattAmplUAA}
\begin{array}{rl}
 \mathcal{M}_{AA}^{\text{U}(N_\text{f})}=&\mathcal{M}_{AA}^{\text{SU}(N_\text{f})} + \frac{2M_\pi^4}{F_\pi^4 N_\text{f}}\left(1+\frac{2}{N_\text{f}}\right)B_1(t)-\frac{2M_\pi^4}{F_\pi^4 N_\text{f}^2}B_2(t)  + (t\leftrightarrow u) \\
&- \left(\frac{M_\pi^2}{F_\pi^2}\right)^3\left[K_{AA}+q^2K_{AA}'\right],
\end{array}
\end{equation}
where $K_R$, and $K_R'$ are the leading-$N_\text{c}$ part of products of $\mathcal{O}(N_\text{c})$ LECs, and $B_1(x)$ and $B_2(x)$ are new loop integrals appearing in these theory that correspond to Figs.~\ref{fig:loop1eta} and~\ref{fig:loop2eta}, 
\begin{equation}
\def\arraystretch{1.5}
\begin{array}{rl}
B_1(z)=&\frac{1}{(4\pi)^2}\left\{\frac{1}{M_{\eta'}^2-M_\pi^2}\left(M_{\eta'}^2\log{\frac{M_{\eta'}^2}{\mu^2}}-M_\pi^2\log{\frac{M_\pi^2}{\mu^2}}\right)\right.\\
&\left.+\int_0^1\text{d}x \log\left[\frac{M_\pi^2x+M_{\eta'}^2(1-x)-x(1-x)z}{M_\pi^2x+M_{\eta'}^2(1-x)}\right]\right\},
\end{array}
\end{equation}
\begin{equation}\label{eq:loopint2}
B_2(z)=\frac{1}{4\pi}^2\left[1+\log{\frac{M_{\eta'}^2}{\mu^2}}+J\left(z\frac{M_\pi}{M_{\eta'}^2}\right)\right].
\end{equation}
 This amplitudes can then be projected to $s$-wave with Eq.~(\ref{eq:partialwave}) and the corresponding phase shift computed using Eq.~(\ref{eq:phaseshift}). Note that, in spite of using the same name, LECs take different values in the two theories, which can be related in the limit of large $\eta'$ mass---see Eqs.(~\ref{eq:I2L1Matching}) and~(\ref{eq:AAL1Matching}).

%\textbf{Menciono los efective ranges?}
%\begin{equation}
%\begin{array}{rl}
%K_0=&(256L_5L_8-512L_8^2)_{\mathcal{O}(N_\text{c}^2)}\\
%K_1=&(256L_5L_8-512L_8^2)_{\mathcal{O}(N_\text{c}^2)}
%\end{array}
%\end{equation}

%From this result, the corresponding phase shift can be extracted, and used to compute the energy levels by means of Lüscher's formalism.

%In the case of $s$-wave interactions ($l=0$), the relation to the scattering length

%These are computed using Lüscher's formalism from the $s$-wave phase shift, as shown in Eq.~(\ref{eq:Luscher}). This quantity can be computed from the corresponding partial wave, which is related to the

% In order to use ChPT to predict the energy shifts of two pions in a particular channel from the lattice, one needs to compute their corresponding phase shift within the theory. 

\label{app:ScattAmpl}
\section{Wilson ChPT with a twisted mass}
\label{app:WilsonChPT}

%The lattice is a non-perturbative regularization scheme that allows to numerically study gauge theories in their confining regime. However, it comes with some drawbacks. First, it breaks chiral symmetry, so quark masses are additively renormalized. Second, it introduces discretization effects to physical observables, with are in general proportional to the lattice spacing, $a$.

%There are different options to reduce these errors. One is to introduce a higher dimensional operators---the Symmanzik terms---whose coefficient, $c_\text{sw}$, is nonperturbatively determine to cancel discretization errors at leading order~\cite{}. Other possibility is to use a twisted mass~\cite{Frezzotti:2000nk}. In the continuum, this is just a chiral rotation of the quark fields which doesn't affect the theory. However, since chiral symmetry is broken in the lattice, this transformation cannot be rotated away and may be used to obtain an $\mathcal{O}(a)$ improvement of some observables. In this work, we have used a non-zero value of $c_\text{sw}$ and compared the results for Wilson and twisted-mass regularizations of the valence sector.

In this work, we have employed two methods to reduce discretization effects. First, we included a Symanzik term in the action~\cite{SHEIKHOLESLAMI, Luscher:1996sc, Luscher:1996ug}, and second, we employed a twisted-mass in the valence sector. In the continuum, the latter is just a chiral rotation of the quark fields, but since chiral symmetry is broken in the lattice, this transformation cannot be rotated away and may be used to obtain an $\mathcal{O}(a)$ improvement of some observables. 

Discretization errors induced by the chosen setup in mesonic observables can be studied using a modification of ChPT that includes the effects of a finite lattice spacing and a twisted mass (known as twisted-mass Wilson ChPT). The theory is written in terms of a SU($N_\text{f})$-valued matrix, $\Sigma$, and two spurion fields, $\chi$ and $A$, all of which transform in the same way under the chiral symmetry, e.g., $\Sigma\rightarrow L\Sigma R^\dagger$, with $L\in \text{SU}(N_\text{f})_\text{L}$ and $R\in \text{SU}(N_\text{f})_\text{R}$. $\Sigma$ is related its usual ChPT counterpart via a chiral rotation, while the spurions include the effects of the twisted mass and the lattice spacing, and are set to constant values,
\begin{equation}
\chi =2B_0(m\mathbb{1}+i\mu T),\quad\quad\quad A=2W_0 a\mathbb{1}\equiv \hat{a}\mathbb{1},
\end{equation} 
with $T$ the twist matrix and $B_0$, $W_0$ two LECs. The chiral Lagrangian then contains one new operator at $\mathcal{O}(p^2)$, and 8 more at $\mathcal{O}(p^4)$, all with their corresponding LEC. Finally, the lattice spacing must be included in the counting scheme. One in general uses
\begin{equation}
\mathcal{O}(m_q)\sim\mathcal{O}(M_\pi^2)\sim\mathcal{O}(k^2)\sim\mathcal{O}(a),
\end{equation}
although $\mathcal{O}(a^2)$ effects are often large and need to be promoted to lower orders.

This theory has been studied in the literature, mainly for $N_\text{f}=2$~\cite{Munster:2004wt, Sharpe:2004ny, Buchoff:2008hh}. In this appendix we review the basis of this extension of ChPT and generalize some results for $N_\text{f}=4$   up to $\mathcal{O}(a^2)$ using $T=\text{diag}(+1,-1,-1,1)$. Note most of the references work in Euclidean signature, while we opt for Minkoski spacetime with the mostly minus choice of the metric.

\subsection{Lagrangian of the theory}

The $\mathcal{O}(p^4)$ Lagrangian of this theory reads
\begin{equation}
\def\arraystretch{1.3}
\begin{array}{rl}
\mathcal{L}  = &  \frac{F^2}{4}\Tr[\partial_\mu\Sigma\partial^\mu\Sigma^\dagger]+\frac{F^2}{4}\Tr[\chi\Sigma^\dagger+\chi^\dagger\Sigma]+\frac{F^2}{4}\Tr[A\Sigma^\dagger+A^\dagger\Sigma]  \\
& +L_0\Tr[\partial_\mu\Sigma\partial_\nu\Sigma\partial^\mu\Sigma^\dagger \partial^\nu\Sigma^\dagger]+L_1\Tr[\partial_\mu\Sigma\partial^\mu\Sigma]^2 \\
&+L_2\Tr[\partial_\mu\Sigma\partial_\nu\Sigma^\dagger] \Tr[\partial^\mu\Sigma\partial^\nu\Sigma^\dagger] +L_3 \Tr[(\partial_\mu\Sigma\partial^\mu\Sigma^\dagger)^2] \\
&+L_4 \Tr[\partial_\mu\Sigma\partial^\mu\Sigma^\dagger]\Tr[\chi\Sigma^\dagger+\chi^\dagger\Sigma] +W_4 \Tr[\partial_\mu\Sigma\partial^\mu\Sigma^\dagger]\Tr[A\Sigma^\dagger+A^\dagger\Sigma]\\

& +L_5 \Tr[\partial_\mu\Sigma\partial^\mu\Sigma^\dagger (\chi\Sigma^\dagger+\chi^\dagger\Sigma)] +W_5 \Tr[\partial_\mu\Sigma\partial^\mu\Sigma^\dagger (A\Sigma^\dagger+A^\dagger\Sigma)] \\
&+L_6\Tr[\chi\Sigma^\dagger+\chi^\dagger\Sigma]^2 +W_6 \Tr[\chi\Sigma^\dagger+\chi^\dagger\Sigma] \Tr[A\Sigma^\dagger+A^\dagger\Sigma] + \tilde{W}_6 \Tr[A\Sigma^\dagger+A^\dagger\Sigma]^2 \\
&+ L_7 \Tr[\chi\Sigma^\dagger-\chi^\dagger\Sigma]^2 + W_7 \Tr[\chi\Sigma^\dagger-\chi^\dagger\Sigma]\Tr[A\Sigma^\dagger-A^\dagger\Sigma] +\tilde{W}_7 \Tr[A\Sigma^\dagger-A^\dagger\Sigma]^2\\
&+ L_8 \Tr[\chi\Sigma^\dagger\chi\Sigma^\dagger+\chi^\dagger\Sigma\chi^\dagger\Sigma]  +W_8 \Tr[\chi\Sigma^\dagger A\Sigma^\dagger+\chi^\dagger\Sigma A^\dagger\Sigma] \\
&+\tilde{W}_8 \Tr[A\Sigma^\dagger A\Sigma^\dagger+A^\dagger\Sigma A^\dagger\Sigma],
\end{array}
\end{equation}
with $W_i$, $\tilde{W}_i$ new LECs that need to be determined numerically. The ones which are relevant in this work have a defined large $N_\text{c}$ scaling,
\begin{equation}
\mathcal{O}(N_\text{c}):W_5, W_8, \tilde{W}_8, \quad\quad\quad \mathcal{O}(1): W_4, W_6, \tilde{W}_6.
\end{equation}
In order to work with this theory, one first redefines the spurion field,
\begin{equation}
\chi'=\chi+A=2B_0(m\mathbb{1} + i\mu T) + W_0a=M^2\text{e}^{i \omega_0 T}
\end{equation}
with $M$ the bare pion mass and $\omega_0$ the twist angle at LO, as well as the new LECs,
\begin{equation}
\begin{array}{rl}
W'_{4,5}&=W_{4,5}-L_{4,5},\\
W'_{6,7,8}&=W_{6,7,8}-2L_{6,7,8},\\
\tilde{W}'_{6,7,8}&=\tilde{W}_{6,7,8}+L_{6,7,8}.
\end{array}
\end{equation}
One then expands around the physical vacuum,
\begin{equation}
\Sigma=\text{e}^{i\omega T/2} \text{e}^{i\phi/F} \text{e}^{i\omega T/2},
\end{equation}
where $F$ is the bare pion decay constant, $\phi$ is the usual pion matrix, and $\omega=\omega_0+\varepsilon$ is the non-perturbative twist angle. At NLO, we can determine $\varepsilon$ by requiring the one-point function to vanish,
\begin{equation}
\varepsilon=-\frac{8\hat{a}s}{F_\pi^2}\left[(4W_6'+W_8')+\frac{2\hat{a}c}{M_\pi^2}(4\tilde{W}_6'+\tilde{W}_8')\right],
\end{equation}
where we have abbreviated $s= \sin\omega$ and $c=\cos\omega$.

\subsection{Discretization effects on physical observables}

We are now in position to study discretization effect for some mesonic observables of interest. %Regarding decay constants, field renormalizations and masses, one first need to expand $\mathcal{L}$ to $\mathcal{O}(\phi^2)$, \textbf{igual no hace falta poner esto}
%\begin{equation}
%\begin{array}{rl}
%\mathcal{L}_{\phi^2}= & \mathcal{L}_{\phi^2}^{\text{cont}}+\frac{2 c\hat{a}}{F^2}(W_4'N_\text{f}+W_5')\Tr[\partial_\mu\phi\partial^\mu\phi]-\frac{2\tilde{W}_8'\hat{a}^2s^2}{F^2}\Tr[T\phi T\phi]-\frac{4\tilde{W}_6's^2\hat{a}^2}{F^2}\Tr[T\phi]^2 \\
%& +\frac{2\hat{a}}{F^2}\left(2W_6'M^2cN_\text{f}+2\tilde{W}_6'c^2\hat{a}N_\text{f}+2W_8'M^2c+2\tilde{W}_8'\hat{a}c^2-\tilde{W}_8'\hat{a}s^2\right)\Tr[\phi^2].
%\end{array}
%\end{equation}
 The lattice corrections to the continuum decay constant, $F_{\pi}^{\text{cont}}$, and field renormalization, $Z^{\text{cont}}$,  are, respectively,
\begin{equation}\label{eq:DecayCorr}
F_\pi=F_{\pi}^{\text{cont}}+\frac{4\hat{a}c}{F_\pi}(4W_4'+W_5'),
\end{equation}
\begin{equation}\label{eq:ZCorr}
Z=Z^{\text{cont}}-\frac{8\hat{a}c}{F_\pi^2}(4W_4'+W_5'),
\end{equation}
all of which vanish at maximal twist. The effects on the pion mass are a bit more complicated, since distinct mesons get different corrections to its continuum value, $M_{\pi}^{\text{cont}}$. For charged mesons one finds
\begin{equation}\label{eq:PiMassCorr}
M_\pm^2=(M_{\pi}^{\text{cont}})^2+\frac{16\hat{a}M_\pi^2c}{F_\pi^2}\left[(4W_6' +W_8')+\frac{c\hat{a}}{M^2}(4\tilde{W}_6' +\tilde{W}_8')\right]-\frac{8M_\pi^2\hat{a}c}{F_\pi^2}(4W_4'+W_5'),
\end{equation}
which also vanishes at maximal twist. However, neutral multiplet mesons get further $\mathcal{O}(a^2)$ corrections that do not cancel in such limit and even lead to a non-diagonal mass matrix for flavorless ones,
\begin{equation}
M_{K^0,D^0}^2=M_\pm^2-\frac{16\tilde{W}_8's^2\hat{a}^2}{F_\pi^2},
\end{equation}
\begin{equation}
\mathbb{M}[\pi^0,\eta^0,\eta_c]=M_{K^0,D^0}^2\mathbb{1}_{3\times 3}-\frac{32\tilde{W}_6' s^2 \hat{a}^2}{3F_\pi^2}\begin{pmatrix}
3 & \sqrt{3} & -\sqrt{6} \\
\sqrt{3} & 1 & -\sqrt{2} \\
-\sqrt{6} & -\sqrt{2} & 2 \\
\end{pmatrix}.
\end{equation}
These corrections have been discussed in the literature to be as large as $\sim30\%$~\cite{Buchoff:2008hh}.

Next, one can consider NLO corrections to the continuum scattering amplitudes in the $SS$ and $AA$ channels, $\mathcal{M}_{SS}^{\text{cont}}$ and $\mathcal{M}_{AA}^{\text{cont}}$,, respectively. These are generated by new four-point vertices proportional to the new LECs, and by corrections to the LO result originated from Eqs.~(\ref{eq:DecayCorr})-(\ref{eq:PiMassCorr}),
\begin{equation}
\mathcal{M}_{SS}=\mathcal{M}_{SS}^{\text{cont}}+\frac{8\hat{a}cM_\pi^2}{F_\pi^4}\left[-2W_4's-W_5's + 8W_6' +4 W_8' + \frac{4\hat{a}c}{M_\pi^2}(2\tilde{W}_6'+\tilde{W}_8')\right],
\end{equation} 
\begin{equation}
\mathcal{M}_{AA}=\mathcal{M}_{AA}^{\text{cont}}+\frac{8\hat{a}cM_\pi^2}{F_\pi^4}\left[-2W_4's+W_5's + 8W_6' -4 W_8' + \frac{4\hat{a}c}{M_\pi^2}(2\tilde{W}_6'-\tilde{W}_8')\right],
\end{equation} 
which are consistent with Ref.~\cite{Buchoff:2008hh}. Note that, as initially expected, all these corrections vanish at maximal twist, meaning that the usage of a properly tuned twisted mass provides automatic $\mathcal{O}(a)$ improvement for these observables. However, this also implies that in order to understand the observed discretization effects for this regularization, we need to go to higher order. For Wilson fermions, on the other hand, $\mathcal{O}(a)$ improvement depends on the correct choice of $c_\text{sw}$, and can only be empirically tested.

\begin{figure}[b!]
\centering
\includegraphics[scale=0.29]{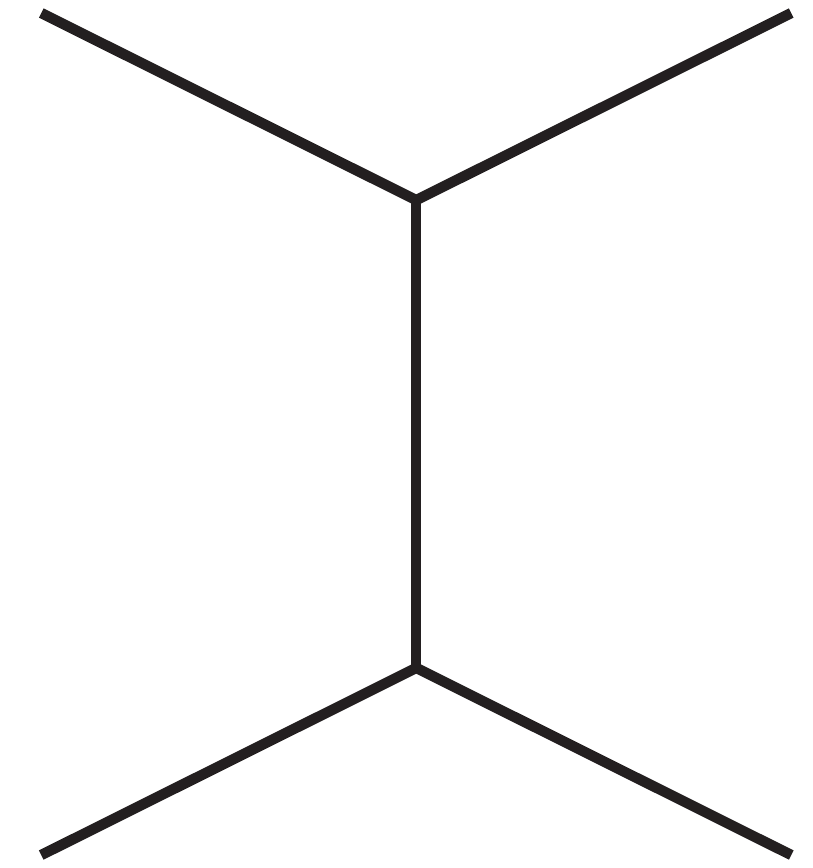} 
\caption{~Additional Feynman diagram required to study $\mathcal{O}(a^2)$ discretization effects in the $AA$ channel for twisted-mass fermions.}\label{fig:DiagE}
\end{figure}

At non-zero twist, further corrections arise from new three-leg vertices that lead to the diagram depicted in Fig.~\ref{fig:DiagE} also contributing to the scattering amplitudes for two channels. For $\omega\neq\pi/2$, this originates $\mathcal{O}(a^4/M_\pi^4)$ and $\mathcal{O}(a^3/M_\pi^2)$ corrections  compared to the leading result. However, both vanish at maximal twist, and only $\mathcal{O}(a^2)$ effects remain
\begin{equation}
\begin{array}{rl}
\mathcal{M}_{SS}^{\omega=\pi/2}=&\mathcal{M}_{SS}^{\text{cont}}+\frac{32\hat{a}^2M_\pi^2}{F_\pi^6}\frac{1}{t-1}\left[8W_4'^2+4W_4'W_5'+W_5'^2-16W_4'W_6'-4W_5'W_6'+8W_6'^2\right.\\
&-4W_4'W_8'-2W_5'W_8'+4W_6'W_8'+W_8'^2+t(-8W_4'^2-4W_4'W_5'-W_5'^2+8W_4'W_6'\\
&+2W_5'W_6'+2W_4'W_8'+W_5'W_8')+\left.\frac{t^2}{4}(8W_4'^2+4W_4'W_5'+W_5'^2)\right]+(t\leftrightarrow u),
\end{array}
\end{equation} 
\begin{equation}
\begin{array}{rl}
\mathcal{M}_{AA}^{\omega=\pi/2}=&\mathcal{M}_{AA}^{\text{cont}}+\frac{32\hat{a}^2M_\pi^2}{F_\pi^6}\frac{1}{t-1}\left[8W_4'^2+4W_4'W_5'-W_5'^2-16W_4'W_6'-4W_5'W_6'+8W_6'^2\right.\\
&-4W_4'W_8'+2W_5'W_8'+4W_6'W_8'-W_8'^2+t(-8W_4'^2-4W_4'W_5'+W_5'^2+8W_4'W_6'\\
&+2W_5'W_6'+2W_4'W_8'-W_5'W_8')+\left.\frac{t^2}{4}(8W_4'^2+4W_4'W_5'-W_5'^2)\right]+(t\leftrightarrow u),
\end{array}
\end{equation} 
where $t$ and $u$ are the usual Mandelstam variables normalized by the pion mass. If we focus on the corrections near threshold and consider only the leading $N_\text{c}$ terms, we get the result explained in Eq.~(\ref{eq:kcotDiscretization}).

Two comments are in place. First, these leading-$N_\text{c}$ corrections should have the same size for both the $SS$ and $AA$ channels. Since discretization effects are not observed in the former, one could argue that the leading corrections are suppressed. We have tried several parametrizations when matching to ChPT and concluded that the chosen one in Eq.~(\ref{eq:kcotDiscretization}) provides the best explanation of the lattice results. Second, through this appendix we have worked in a generalization of SU($N_\text{f}$) ChPT. The study of a similar generalization of the U($N_\text{f}$) theory is out of the scope of this paper, and is left for future work.

%quenched and eta'

\label{app:WChPT}

\clearpage
\bibliographystyle{JHEP}      
\bibliography{bibtexref.bib}

\end{document}